\documentclass[journal]{IEEEtran}
\ifCLASSINFOpdf
\else
\fi
\usepackage{graphicx}
\usepackage{amsmath}
\usepackage{amssymb}
\usepackage{cite}
\usepackage{subfigure}
\usepackage{booktabs}
\usepackage{color}
\usepackage{balance}
\usepackage{subfig}
\usepackage{subfloat}

\usepackage{epstopdf}
\usepackage{enumerate}
\usepackage{amsthm}
\usepackage{algorithm}
\usepackage{algorithmic}

\newtheorem{definition}{Definition} 
\newtheorem{assumption}{Assumption} 
\newtheorem{theorem}{Theorem}
\newtheorem{lemma}[theorem]{Lemma}

\begin{document}
%
\title{Linearly Constrained Smoothing Group Sparsity Solvers in Off-grid Model}
%
%
%

\author{Cheng-Yu~Hung,~\IEEEmembership{Member,~IEEE,}
        and~Mostafa~Kaveh,~\IEEEmembership{Life~Fellow,~IEEE}
\thanks{C. Y. Hung was with the Department
of Electrical and Computer Engineering, University of Minnesota - Twin Cities, Minneapolis,
MN, 55455 USA e-mail: hungx086@umn.edu.}
\thanks{M. Kaveh is with University of Minnesota.}
}

\maketitle

\begin{abstract}
In compressed sensing, the sensing matrix is assumed perfectly known. However, there exists perturbation in the sensing matrix in reality due to sensor offsets or noise disturbance. Directions-of-arrival (DoA) estimation with off-grid effect satisfies this situation, and can be formulated into a (non)convex optimization problem with linear inequalities constraints, which can be solved by the interior point method (using the CVX tools), but at a large computational cost. In this work, in order to design efficient algorithms, we consider various alternative formulations, such as unconstrained formulation, primal-dual formulation, or conic formulation to develop group-sparsity promoted solvers. First, the consensus alternating direction method of multipliers (C-ADMM) is applied. Then, iterative algorithms for the BPDN formulation is proposed by combining the Nesterov smoothing technique with accelerated proximal gradient method, and the convergence analysis of the method is conducted as well. 
 We also developed a variant of EGT (Excessive Gap Technique)-based primal-dual method to systematically reduce the smoothing parameter sequentially. Finally, we propose algorithms for quadratically constrained $\ell_2$-$\ell_1$ mixed norm minimization problem by using the smoothed dual conic optimization (SDCO) and continuation technique.
The performance of accuracy and convergence  for all the proposed methods are demonstrated in the numerical simulations.

\end{abstract}


%
\IEEEpeerreviewmaketitle

\section{Introduction}
%
%
%
%
\IEEEPARstart{I}{n} compressed sensing \cite{donoho2006compressed,candes2006robust}, an underdetermined linear system  is considered
\begin{align} 
{\bf y}={\bf A}{\bf s}+{\bf n},
\end{align}
where ${\bf y} \in {\mathbb C}^{M\times 1}$ is an observation measurement vector, ${\bf A} \in {\mathbb C}^{M\times N} (M\ll N)$ is a known dictionary matrix,  ${\bf n}\in {\mathbb C}^{M\times 1}$ is a measurement error or additive noise vector, and ${\bf s}\in {\mathbb C}^{N\times 1}$ is a $K$-sparse signal vector of interest. There are only $K$ nonzero entries in $\bf s$, and $K\ll N$.
As long as the dictionary matrix $\bf A$ meets the requirement of the Restricted Isometry Property (RIP) \cite{candes2006robust,duarte2011structured,eldar2012compressed}, the sparse vector $\bf s$ can be reconstructed even with a few measurements by  many solvers, such as group Lasso (least absolute shrinkage and selection operator) \cite{yuan2006model}, basis pursuit denoising (BPDN) \cite{chen2001atomic}, or Dantzig selector \cite{candes2007dantzig}. The performance analysis and computable performance bounds of these sparse recovery solvers are conducted in \cite{tang2010performance,tang2015computable}. 
However,  the dictionary matrix $\bf A$ may not be known perfectly due to certain noise or modeling perturbations. In \cite{chi2011sensitivity}, the sensitivity of basis mismatch in the dictionary matrix is analyzed.
For instance, the compressed sensing approach for DoA estimation may assume a known dictionary formed from the array responses at a grid of candidate directions \cite{malioutov2005sparse}. In practice, however, the DoAs are most likely not to locate on the model grid, leading to the now well-known off-grid DoA estimation problem, for which a number of model approximations and solutions have been proposed, for example \cite{zhu2011sparsity,yang2012robustly,zheng2013sparse,tan2014joint,tang2013compressed,jagannath2013block,lin2014super}.
A commonly-used observation for off-grid DoAs follows the noisy structured perturbation model given by:
\begin{align}\label{cs_off}
{\bf y}=({\bf A}+{\bf B}\Gamma){\bf s}+{\bf n},
\end{align}
where ${\bf A} \in {\mathbb C}^{M\times N}$ is known, and ${\bf B}\in {\mathbb C}^{M\times N}$ is known as part of the off-grid approximation. $\Gamma=diag(\boldsymbol{\beta}) \in {\mathbb R}^{N\times N}$, and ${\boldsymbol{\beta}}=[\beta_1,\dots,\beta_N]^T$ is denoted as the unknown coefficient vector for the approximation. ${\bf s} \in {\mathbb R}^{N\times 1}$ is the sparse vector associated with grid points nearest the true DoAs.
 Equation (\ref{cs_off}) can be solved by formulating a sparsity promoting constrained nonconvex minimization problem to estimate $\bf s$ and $\boldsymbol{\beta}$ sequentially by the alternating method \cite{zhu2011sparsity,yang2012robustly}, but with slow convergence. The alternating direction method of multipliers (ADMM)\cite{boyd2011distributed} is a very popular method, which can be applied to solve this problem.

Furthermore, many inverse problems in signal processing, data mining, or statistical machine learning can be cast as a composite optimization problem, which involves the minimization of a sum of differentiable functions and nonsmooth ones. The off-grid DoA estimation problem of (\ref{cs_off}) can be formulated into this type of composite form. Subgradient algorithms \cite{ben1989smoothing} are developed to deal with nonsmooth optimization problems but with very slow convergence rate.
Instead of using subgradient methods, we attempt to design algorithms for solving nonsmooth optimization (NSO) problems efficiently by using a sequence of approximate smoothing problems to substitute for the original ones. The core of the techniques considered is to make the nondifferentiable functions smooth without introducing substantial approximate errors caused by the smoothing process.
Several different smoothing techniques have been proposed to solve NSO problems \cite{nesterov2005smooth,bertsekas1975nondifferentiable,moreau1965proximite}. A primal-dual symmetric method derived from the excessive gap condition for nonsmooth convex optimization is proposed in \cite{nesterov2005excessive}.
 In \cite{tan2014joint}, the nondifferentiable function, which is approximated by the Moreau envelope function \cite{moreau1965proximite}, is used in the column-wise mismatch problem.  In \cite{chen2012smoothing}, the overlapping group-lasso  penalty is smoothed by the Nesterov smoothing technique \cite{nesterov2005smooth}. A unified framework of smoothing approximation with fast gradient schemes is proposed in \cite{beck2012smoothing}. In \cite{tran2015adaptive}, an adaptive Nesterov-based smoothing method is developed to dynamically choose the smoothing parameter at each iteration of the update. 
 In \cite{komodakis2015playing}, a number of primal-dual iterative approaches for solving large-scale  nonsmooth optimization problems, such as the M+LFBF (Monotone+Lipschitz Forward Backward Forward) algorithm, are reviewed. In \cite{shor2012minimization,nesterov2013introductory}, subgradient methods are proposed, but their complexity cannot be better than than ${\mathcal O}(\frac{1}{\sqrt k})$ where $k$ is the number of iterations. Alternatively, smoothing as presented in \cite{nesterov2005smooth} can be applied to mitigate non-smoothness of the objective function. In \cite{orabona2012prisma}, a proximal iterative smoothing algorithm was proposed to solve convex nonsmooth optimization problems.

In this work, an unconstrained off-grid DoA estimator is first discussed. It consists of one differentiable function and two nonsmooth ones, which are a regularized group-sparsity penalty and an indicator function. First, the consensus ADMM (C-ADMM) \cite{boyd2011distributed} is applied to solve this unconstrained optimization problem by using a common global variable which makes all the local variables of objective functions equal, but it can be very slow to converge to high accuracy. In order to have a low reconstruction error of DoA estimation quickly, the Nesterov smoothing methodology \cite{nesterov2005smooth,chen2012smoothing} is used to reformulate the group-sparsity penalty into a "max"-structure function, and then smoothing it by adding a strongly convex term. We propose two reformulations for the group-sparsity penalty since $\ell_{2}$-$\ell_{1}$ mixed norm has a two-layer norm structure. Then, the accelerated proximal gradient \cite{parikh2014proximal} method is used on the smoothed optimization case. Note that our first proposed Nesterov smoothing method is equivalent to the one in \cite{tan2014joint}, as can be deduced from the results of \cite{orabona2012prisma}. 
The second Nesterov smoothing method is proposed by use of the property of dual of $\ell_1$ norm. 
It's noted that the fixed smoothing parameter  has to be chosen empirically in this method. However, \cite{becker2011nesta} shows that the accuracy performance increases when the smoothing parameter decreases. Thus, by the excess gap technique (EGT) \cite{nesterov2005excessive},  in order to reduce the smoothing parameter sequentially, we developed a variant of EGT-based primal-dual method, in which a surrogate of cost function is introduced. Furthermore, inspired by \cite{becker2011templates,becker2011nesta}, a variant of conic formulation for quadratically constrained $\ell_{2}$-$\ell_{1}$ mixed norm minimization with linear ineuqalities is proposed, and solved by using the smoothed dual conic optimization and continuation technique.
The accuracy, and convergence of performance for the proposed methods are demonstrated, and compared with the interior point method (CVX) \cite{SBoyd2004}, MUSIC \cite{schmidt1986multiple}, M+LFBF \cite{komodakis2015playing}, and CRLB \cite{stoica1990performance}.

This paper is organized as follows. In Section II, Some mathematical preliminaries, and the off-grid DoA model with its C-ADMM solver are introduced. In Section III, the Nesterov smoothing technique is employed to reformulate the group-sparsity penalty in two ways. Then, accelerated smoothing proximal gradient (ASPG) is used to solve the reformulated optimization problems. The convergence behavior is analyzed as well. In Section IV, the EGT-based approach is utilized to provide a systematic way to reduce the smoothing parameter. Finally, in Section V, the smoothing technique is applied in the conic formulation on the off-grid DoA estimation. Section VI presents numerical results to verify the performance in terms of DoA resolution ability, estimation accuracy, and convergence behavior.

{\it Notation}: Throughout the paper, vectors and matrices are represented by boldface lowercase and uppercase letters, respectively. $E(\cdot)$ denotes the expectation operator. For any given matrix $\bf X$, ${\bf X}^H$ denotes the Hermitian transpose matrix, and $vec({\bf X})$ is the vectorization operator of the matrix. $diag({\bf x})$ represents a diagonal square matrix with the elements of vector $\bf x$ on the diagonal.
$\odot$ denotes the Hadamard product. $\otimes$ denotes the Kronecker product. For any two vectors ${\bf x},{\bf y}$, $ ({\bf x},{\bf y})$ is denoted as a new vector in which $\bf x$ is stacked by $\bf y$, and $\langle {\bf x}, {\bf y} \rangle$ means the inner product.   $Proj_{\mathcal X}({\bf x})$ denotes the projection operator of projecting a vector $\bf x$ onto a space ${\mathcal X}$.

\section{Preliminaries, DoA Model with Structured Perturbations, and C-ADMM solver}
\subsection{Preliminaries}

Consider the following unconstrained separable convex optimization problem \cite{johnson2013accelerating}:
\begin{align} \label{manyfuns}
\min_{{\bf x} \in {\mathbb R}^n} F({\bf x}),~~~  F({\bf x}):= \sum_{i=1}^n f_i({\bf x}),  
\end{align}  
where $\{ f_i({\bf x}),\cdots,f_n({\bf x}) \}$ is a sequence of convex functions from ${\mathbb R}^n$ to $\mathbb R$.

In this paper, specifically, an unconstrained convex optimization problem is considered:
\begin{align} \label{3funs}
\min_{{\bf x} \in {\mathbb R}^n} F({\bf x})= \{ f({\bf x})+h({\bf x})+i({\bf x}) \} ,  
\end{align}     
that satisfy the following assumptions, and definitions:
\begin{assumption}
\end{assumption}
\begin{enumerate}[(i)]
\item $f:{\mathbb R}^n\rightarrow {\mathbb R}\cup \{+ \infty\}$ is a proper, closed, convex and continuously differentiable function. Its gradient is Lipschitz continuous with parameter $L_f$. 
\item $h:{\mathbb R}^n\rightarrow {\mathbb R}\cup \{+ \infty\}$ is a proper, closed, and convex $\rho_h$-Lipschitz continuous function. It is not necessarily differentiable.
\item $i:{\mathbb R}^n\rightarrow {\mathbb R}\cup \{+ \infty\}$ is a proper, lower semicontinuous, and convex function but possibly nonsmooth. For instance, the indicator function of a closed set is lower semi-continuous.
\end{enumerate}
\begin{definition}[Lipschitz Continuous]
A function $f:{\mathbb R}^n\rightarrow {\mathbb R}$ is $\rho$-Lipschitz continuous if there exits $ \rho>0$ such that $| f({\bf x})- f({\bf y})| \leq {\rho} \| \bf x - y \|$, $\forall {\bf x, y} \in \mathbb R^n$.
\end{definition}
\begin{definition}[Lipschitz Continuous Gradient]
The gradient of a differentiable convex function $f:{\mathbb R}^n\rightarrow {\mathbb R}$ is Lipschitz continuous with parameter $L>0$ if $\| {\nabla f({\bf x})}- {\nabla f({\bf y})} \| \leq L \| \bf x - y \|$, $\forall {\bf x, y} \in \mathbb R^n$.
\end{definition}
\begin{definition}[Strongly Convex]\label{def3}
The function $f:{\mathcal X}\rightarrow {\mathbb R}$ is $\sigma$-strongly convex on a closed convex set $\mathcal X$ with parameter $\sigma>0$ if $ f({\bf y}) \geq  f({\bf x}) +\nabla f({\bf x})^T({\bf y - x}) +\frac{\sigma}{2} \| \bf y - x \|_2^2$, $\forall {\bf x, y} \in \mathcal X$.
\end{definition}

In the next subsection, we will show that the DoA estimation problem with structured perturbations can be reformulated into the form of (\ref{3funs}).

\subsection{DoA Model with Structured Perturbations}
Consider an array of $M$ sensors and suppose that there are $K$ far-field narrowband sources impinging on the array from angles $\theta_1, \dots, \theta_K$. The measurement model, and its covariance are described by
\begin{align}\label{single-snap}
&{\bf v}(t)=\sum_{k=1}^K {\tilde s}_k(t){\bf a}(\theta_k)+{\bf n}(t)={\tilde{ \bf A}(\boldsymbol{\theta})}{\tilde{\bf s}}(t)+{\bf n}(t), \\
&{\bf {\  R}_{v}}= E[{\bf vv}^H]  = \sum_{k=1}^{K}\sigma^2_k{\bf a}(\theta_k){\bf a}(\theta_k)^H+\sigma_n^2{\bf I},
\end{align}
where
\begin{itemize}
\item ${\bf v}(t) \in {\mathbb C}^{M\times 1}$ is the observation vector.
\item ${\tilde s}_k(t)$ is the $k$-th received signal with power $\sigma^2_k$.
\item ${\bf a}(\theta_k)$ denotes the steering vector for direction $\theta_k$ with $m$-th entry $e^{-j2\pi \frac{d_m}{\lambda} sin\theta_k}$, where $\lambda$ is wavelength. $\tilde{\bf A}(\boldsymbol{\theta})=[{\bf a}(\theta_1),\dots,{\bf a}(\theta_K)]$.
\end{itemize}


In compressed sensing, $\boldsymbol{\phi}=[\phi_1,\dots,\phi_N]$ is defined as  uniformly discretized grid atoms for the dictionary matrix. The off-grid DoA is denoted by $\beta_i=\theta_k-\phi_i$ if $\phi_i$ is closest to $\theta_k,\forall k$; otherwise, $\beta_i =0$. We assume that $0\leq |\beta_i|\leq r$ and $r=\frac{|{\phi_i-\phi_{i+1}}|}{2}$.

By using Taylor series, the first-order approximate measurement model \cite{hung2014directions} is
\begin{align}\label{taylor_appr} 
&\tilde{{\bf v}}(t)=(\tilde{\bf A}(\boldsymbol{\phi})+\tilde{{\bf B}}\Gamma){\bar{\bf s}}(t)+{\bf n}(t),
\end{align}
where $\tilde{{\bf B}}=[\frac{\partial{\bf a}(\phi_1)}{\partial \phi_1},\dots,\frac{\partial{\bf a}(\phi_N)}{\partial \phi_N}]\in {\mathbb C^{M\times N}}$, ${\boldsymbol{\beta}}=[\beta_1,\dots,\beta_N]^T$, $\Gamma=diag(\boldsymbol{\beta})$, and $\bar {\bf s}$ is a $\mathbb C^{N\times 1}$ sparse vector. 
$\tilde{\bf A}(\boldsymbol{\phi})=[{\bf a}(\phi_1),\dots,{\bf a}(\phi_N)]$.
By vectorizing the covariance of (\ref{taylor_appr}), we have 
\begin{align}\label{vec_cov} 
{\bf y}&=({\bf A}(\boldsymbol{\phi})+{\bf B}\Gamma){{\bf s}} +\sigma_n {\bf 1}_n \\ \nonumber
&=({\bf A}(\boldsymbol{\phi}){\bf  s}+{\bf B}{\bf p}) +\sigma_n {\bf 1}_n=[{\bf A}(\boldsymbol{\phi}) , {\bf B}]{\bf x} +\sigma_n {\bf 1}_n,  
\end{align}
where
\begin{itemize}
\item ${\bf y}=vec({\bf R_{\tilde v}})$.
\item ${\bf A}(\boldsymbol{\phi})=[{\bf a}(\phi_1)^H \otimes{\bf a}(\phi_1),\dots,{\bf a}(\phi_N)^H \otimes{\bf a}(\phi_N)]\in  {\mathbb C^{M^2\times N}}$.
\item ${{\bf B}}=[\frac{\partial{\bf a}(\phi_1)}{\partial \phi_1}\otimes \frac{\partial{\bf a}(\phi_1)}{\partial \phi_1},\dots,\frac{\partial{\bf a}(\phi_N)}{\partial \phi_N}\otimes \frac{\partial{\bf a}(\phi_N)}{\partial \phi_N}]\in {\mathbb C^{M^2\times N}}$.
\item ${\bf s}$ is a $\mathbb R^{N\times 1}$ sparse vector with $K$ nonzero terms $\sigma_k^2$'s.
\end{itemize}
${\bf 1}_n=[e_1^T,\dots,e_M^T]^T$ where $e_i \in {\mathbb R}^{M\times 1}$ is an all-zero vector except with 1 at $i$-th entry. ${\bf x}=[{\bf s}^T,{\bf p}^T]^T\in {\mathbb R^{2N\times 1}}$, and ${\bf p}={\boldsymbol \beta}\odot {\bf s}$. Let ${\bf G}=[{\bf A}(\boldsymbol{\phi}) , {\bf B}]$ be a fat matrix for the following sections. 
Note that if $r$ is less than or equal to $0.5$, then ${\bf s} \gg {\bf p}$ since the value of $\beta_k$ is much smaller than $\sigma^2_k$ at mild SNRs.

Since ${\bf s, p}$ have the same sparsity pattern (non-zero entries), we can solve (\ref{vec_cov}) over a closed convex set $\mathcal X$ by the group Lasso :
\begin{equation}\label{glasso_std}
\begin{aligned}
& \underset{{\bf x}\in {\mathcal X}}{\text{arg min}}~~ \frac{1}{2}||{\bf y}-{{\bf G}}{\bf x}||_2^2 + \eta||{\bf x}||_{2,1},  \\
& \text{s.t.}~~ {\mathcal X}:=\{ {\bf x}=[{\bf s}^T,{\bf p}^T]^T : {\bf s}\geq0 , -r{\bf s}\leq {\bf p} \leq r{\bf s}\}.
\end{aligned}
\end{equation}
where $\eta >0$ is a regularization parameter, and $r$ is defined previously. 
Because the constraint set ${\mathcal X}$ is a linear inequalities constraint, we can transform it into an unconstrained one by using an indicator function, which is also known as the basis pursuit denoising problem (BPDN) formulation:
\begin{align}\label{glasso_variant}
 \arg\min_{{\bf x}\in {\mathbb R}^{2N\times 1}} F({\bf x})=\{  \frac{1}{2}||{\bf y}-{{\bf G}}{\bf x}||_2^2 + \eta||{\bf x}||_{2,1}+{\iota}_{\mathcal X}(\bf x)\},
\end{align}
where ${\iota}_{\mathcal X}(\bf x)=0$ if ${\bf x} \in {\mathcal X}$; otherwise, $\infty$. 
Let $f({\bf x}):=\frac{1}{2}||{\bf y}-{{\bf G}}{\bf x}||_2^2$, $h({\bf x}):= \eta||{\bf x}||_{2,1}$, and $i({\bf x}):= {\iota}_{\mathcal X}(\bf x)$ such that (\ref{glasso_variant}) fits the framework of (\ref{3funs}).
Our goal is to solve an optimal solution of problem (\ref{glasso_variant}) efficiently.
However, two nonsmooth functions, $||{\bf x}||_{2,1}$ and ${\iota}_{\mathcal X}(\bf x)$, in the objective makes this problem difficult to solve it.
Thus, the C-ADMM is applied to overcome this situation.

\subsection{Consensus Alternating Direction Method of Multipliers (C-ADMM)}

Let us consider the unconstrained problem (\ref{glasso_variant}).
This problem can be solved by C-ADMM, which uses a consensus global variable ${\bf x}$ and local variables ${\bf z}_i$:  
\begin{align} \nonumber
 &\arg\min_{{\bf x},{\bf z}_i}    \frac{1}{2}||{\bf y}-{{\bf G}}{\bf z}_1||_2^2+\eta||{\bf z}_2||_{2,1}+ {\iota}_{\mathcal X}({\bf z}_3). \\  
& s.t.~~~~ {\bf z}_1 ={\bf x}, ~{\bf z}_2 ={\bf x},~ {\bf z}_3 ={\bf x}
\end{align} We call this a "consensus problem" since the constraint forces all the local variables to be equal. 

C-ADMM of this  problem can be derived from the augmented Lagrangian
\begin{align} 
 L_{\rho}({\bf z, x, u})= \sum_{i=1}^3( f_i({\bf z}_i)+ {\bf u}_i^T({\bf z}_i-{\bf x})+ \frac{\rho}{2} \| {\bf z}_i-{\bf x} \|^2_2) , 
\end{align}
where $ f_1({\bf z}_1)= \frac{1}{2}||{\bf y}-{{\bf G}}{\bf z}_1||_2^2$, $ f_2({\bf z}_2)= \eta||{\bf z}_2||_{2,1}$, $ f_3({\bf z}_2)= {\iota}_{\mathcal X}({\bf z}_3)$, and $\rho$ is a  penalty parameter.
The resulting consensus ADMM is summarized in Algorithm \ref{alg::CADMM}.
\renewcommand{\algorithmicrequire}{\textbf{Input:}}
    \renewcommand{\algorithmicensure}{\textbf{Step k:}}
    \renewcommand{\algorithmicuntil}{\textbf{return}}

\begin{algorithm}[htb]
  \caption{Consensus Alternating Direction Method of Multipliers (C-ADMM)}
  \label{alg::CADMM}
  \begin{algorithmic}[1]
    \REQUIRE
      ${\bf x}^0={\bf 0}$, ${\bf z}_i^0={\bf 0} , \forall i$, ${\bf u}^0=[{\bf u}_1^0, {\bf u}_2^0,{\bf u}_3^0]={\bf 0}$, $\rho=1$

    \ENSURE ($k\geq 0$)   
      \STATE ${\bf z}_1^{k+1}=\arg\min_{{\bf z}_1}  L_{\rho}({\bf z}_1, {\bf x}^k, {\bf u}^k)$ \\
    $\Rightarrow {\bf z}_1^{k+1}=({\bf G}^H{\bf G}+\rho{\bf I})^{-1}({\bf G}^H{\bf y}+\rho{\bf x}^k-{\bf u}_1^k )$	\\
          ${\bf z}_2^{k+1}=\arg\min_{{\bf z}_2}  L_{\rho}({\bf z}_2, {\bf x}^k, {\bf u}^k)$ \\
     $\Rightarrow {\bf z}_2^{k+1}=\frac{{\bf x}^k+{\bf u}_2^k/\rho}{\| {\bf x}^k+{\bf u}_2^k/\rho\|_2} \max(\| {\bf x}^k+{\bf u}_2^k/\rho\|_2-\eta/\rho,0)$	\\
         ${\bf z}_3^{k+1}=\arg\min_{{\bf z}_3}  L_{\rho}({\bf z}_3, {\bf x}^k, {\bf u}^k)$$=Proj_{\mathcal X}({\bf x}^k-\frac{{\bf u}_3^k}{\rho})$	\\
       \STATE ${\bf x}^{k+1}=\arg\min_{{\bf x}}  L_{\rho}({\bf z}^{k+1}, {\bf x}, {\bf u}^k)$$=\frac{\rho(\sum_i{\bf z}_i +\sum_i {\bf u}_i)}{3\rho}$	\\
       \STATE ${\bf u}_1^{k+1}={\bf u}_1^k+ \rho({\bf z}_1^{k+1}-{\bf x}^{k+1})$ \\	
                  ${\bf u}_2^{k+1}={\bf u}_2^k+ \rho({\bf z}_2^{k+1}-{\bf x}^{k+1})$ \\	
                  ${\bf u}_3^{k+1}={\bf u}_3^k+ \rho({\bf z}_3^{k+1}-{\bf x}^{k+1})$ \\	
  \end{algorithmic} 
\end{algorithm}

The convergence of C-ADMM is in terms of the following two assumptions:
\begin{assumption}\label{assump2}
The extended-real-valued function $f_i({\bf z}_i): {\mathbb R}^n\rightarrow {\mathbb R} \cup \{+\infty\} $ are closed, proper, and convex.
\end{assumption}
\begin{assumption}\label{assump3}
The unaugmented Lagragian $L_0$ has a saddle point. Namely, there exists a not necessarily unique solution $({\bf z}^{\ast},{\bf x}^{\ast},{\bf u}^{\ast})$ such that
\begin{align} 
L_0({\bf z}^{\ast},{\bf x}^{\ast},{\bf u}) \leq  L_0({\bf z}^{\ast},{\bf x}^{\ast},{\bf u}^{\ast})  \leq  L_0({\bf z},{\bf x},{\bf u}^{\ast})
\end{align}
\end{assumption}
In \cite{boyd2011distributed}, under assumptions \ref{assump2} and \ref{assump3}, C-ADMM is shown to have its iterations satisfy residual convergence, objective convergence, and dual variable convergence.  The update steps of C-ADMM is summarizes in the Algorithm 1.



\section{The Smoothing Techniques}
In the following sections,  we will show how to deal with problem (\ref{glasso_variant}) by combining the accelerated proximal gradient (APG) algorithm with the Nesterov smoothing technique. We aim to smooth the group-sparsity penalty $h({\bf x})= \eta||{\bf x}||_{2,1}$ so that the APG method can be used. A variant of EGT-based primal-dual method and smoothed dual conic optimization method will be described in the following sections. In order to present the idea more clearly, we introduce the notation $||{\bf x}||_{2,1}=\sum_{g_i \in {\Omega}}  \| {\bf x}_{g_i}\|_2$, where ${\bf x}_{g_i}\in {\mathbb R}^{|g_i|}$ denotes the subvector of ${\bf x} \in {\mathbb R}^{2N\times 1}$ having the same sparse pattern in group $g_i$, where $| \cdot |$ is the cardinality of a set. Each group $g_i$ represents a subset of index set $\{1,\cdots,2N \}$ and is disjoint from the others. Denote $\Omega=\{ g_1,\dots,g_{|\Omega|} \}$ as the set of groups, and $2N=\sum_{i=1}^{|\Omega|}|g_i|$. In our case, $|\Omega|=N$, $|g_i|=2, g_i=\{i,i+N\}, \forall i=1,\cdots,N$, ${\bf x}_{g_i}=[{ x}_i, { x}_{i+N}]^T \in {\mathbb R}^{2}$ where ${ x}_i={  s}_i$ and ${  x}_{i+N}={ p}_i$. Denote ${  x}_i$, ${s}_i$, and ${ p}_i$ as the $i$-th entry of $\bf x, s$, and $\bf p$, respectively.


\subsection{Two Reformulations of Group-sparsity Penalty}

Since $h({\bf x})$ is an $\ell_2$-$\ell_1$ mixed norm with two layers, i.e., the inner is $\ell_2$ norm and the outer is $\ell_1$ norm, we can utilize the dual norm property to reformulate it as a maximization of a linear function over an auxiliary variable with "simple" constraints in two different ways. 

First, inspired by \cite{chen2012smoothing}, by using the convex conjugate function and the fact that the dual norm of  $\ell_2$ norm is  $\ell_2$ norm, $\| {\bf x}_{g_i}\|_2$ has the max-structure as $\max_{\|{\bf u}_{g_i}\|_2\leq1}{\bf u}_{g_i}^T {\bf x}_{g_i}$ where ${\bf u}_{g_i} \in {\mathbb R}^{|g_i|}$ denotes an auxiliary vector. Then, $h({\bf x})$  can be written as
\begin{align}\label{reform_l2} \nonumber
h({\bf x})&=\eta||{\bf x}||_{2,1}= \eta\sum_{g_i \in {\Omega}}  \| {\bf x}_{g_i}\|_2=\sum_{g_i \in {\Omega}}\max_{\|{\bf u}_{g_i}\|_2\leq1}  \eta \langle{\bf x}_{g_i} , {\bf u}_{g_i} \rangle  \\ 
&=\max_{{\bf u} \in {\mathcal U}_{l_2}}  \sum_{g_i \in {\Omega}}  \eta \langle{\bf x}_{g_i} , {\bf u}_{g_i} \rangle  =\max_{{\bf u} \in {\mathcal U}_{l_2}}   \eta  \langle{\bf x} , {\bf u}\rangle , 
\end{align} 
where 
\begin{align}
{\mathcal U}_{l_2}=\{  {\bf u} \in {\mathbb R}^{2N\times 1} : {\|{\bf u}_{g_i}\|_2\leq1}, \forall {g_i}\in \Omega \}
\end{align}
 is the set of vectors in the space of the Cartesian product of $\ell_2$ norm unit ball. In the Nesterov smoothing technique, if a nonsmooth convex function has the max-structure, then we have its corresponding smoothed function  
\begin{align} \label{smooth_fn_l2}
h^{l_2}_{\mu}({\bf x}):=\max_{{\bf u} \in {\mathcal U}_{l_2}} \{  \eta  \langle{\bf x} , {\bf u}\rangle -\mu d_{l_2}({\bf u})  \}
\end{align} 
with a smoothing parameter $\mu >0$, where a $prox$-$function$ $d_{l_2}({\bf u})$  \cite{nesterov2005smooth} is continuous and strongly convex on ${\mathcal U}_{l_2}$ with a strong convexity parameter $\sigma$. Its $prox$-$center$ of $d(\bf u)$ is denoted by ${\bf u}_0=\arg\min_{{\bf u}\in {\mathcal U}_{l_2}}\{ d_{l_2}({\bf u}) \}$. By the definition of strongly convex, $d_{l_2}({\bf u}) \geq \frac{\sigma}{2}\| {\bf u} - {\bf u}_0 \|^2_2$. Since $d_{l_2}({\bf u})$ is strongly convex, $h^{l_2}_{\mu}({\bf x})$ is a smooth and convex function so that its solution is unique and its gradient can be computed easily.


Second, inspired by the fact that the dual norm of $\ell_1$ norm is  $\ell_{\infty}$ norm, $\| {\bf x}\|_1$ has the max-structure as $\max_{\|{\bf u} \|_{\infty}\leq1}{\bf u} ^T {\bf x} $, where ${\bf u}  $ denotes an auxiliary vector. Therefore, we propose a second reformulation. Let us define $\nu_{i}:= \| {\bf x}_{g_i}\|_2$ and ${\boldsymbol \nu}=[\nu_{1},\dots,\nu_{{|\Omega|}}]^T\in {\mathbb R}^{N\times 1}$, and then $h({\bf x})$ can be rewritten as 
\begin{align} 
h({\bf x})&= \eta\sum_{g_i \in {\Omega}}  \| {\bf x}_{g_i}\|_2=\eta \sum_{i=1}^{{|\Omega|}}\nu_{i}=\eta \| \nu \|_1. 
\end{align}
We define a new function ${\bar h}({\boldsymbol \nu})$ as
\begin{align}
{\bar h}({\boldsymbol \nu})=\eta \| \nu \|_1 =\max_{{\bf u} \in {\mathcal U}_{l_1}}   \eta  \langle{\boldsymbol \nu} , {\bf u}\rangle  ,
\end{align} 
where 
\begin{align}
{\mathcal U}_{l_1}=\{  {\bf u} \in {\mathbb R}^{N\times 1} : {\|{\bf u}\|_{\infty} \leq1}  \}
\end{align}
is the set of vectors in the space of  $\ell_{\infty}$ norm unit ball. Since it has the max-structure, we have the corresponding smoothed function of ${\bar h}({\boldsymbol \nu})$ as  
\begin{align} \label{smooth_fn_l1}
h^{l_1}_{\mu}({\boldsymbol \nu}):=\max_{{\bf u} \in {\mathcal U}_{l_1}} \{  \eta  \langle {\boldsymbol \nu} , {\bf u}\rangle -\mu d_{l_1}({\bf u})  \} 
\end{align} 
with a smoothing parameter $\mu >0$. Then, $h^{l_1}_{\mu}({\boldsymbol \nu})$ is also a smooth and convex function if a strongly convex function $d_{l_1}({\bf u})$ is chosen. Note that the dimension of $\bf x$ is twice as many as $\boldsymbol \nu$.

Since both $h^{l_2}_{\mu}({\bf x})$ and $h^{l_1}_{\mu}({\boldsymbol \nu})$ are smooth and convex, their gradients can be formed by the following modified theorem \cite{nesterov2005smooth} 
\begin{theorem}
For any $\mu >0$, the functions $h^{l_2}_{\mu}({\bf x})$ and $h^{l_1}_{\mu}({\boldsymbol \nu})$ are well-defined and continuously differentiable in $\bf x$ and ${\boldsymbol \nu}$, respectively. Moreover, both functions are convex and their gradients:
\begin{align}\label{smoothL2}
\nabla h^{l_2}_{\mu}({\bf x})=\eta {\bf u}^{l_2},~~~~~
\nabla h^{l_1}_{\mu}({\boldsymbol \nu})=\eta {\bf u}^{l_1} 
\end{align} 
are Lipschitz continuous with the same constant $L_{\mu}=\frac{1}{\mu \sigma} $, where ${\bf u}^{l_2} $ and ${\bf u}^{l_1} $ are the optimal solutions to (\ref{smooth_fn_l2}) and (\ref{smooth_fn_l1}), respectively.
\end{theorem}

Suppose that $ \forall {\bf u} \in {\mathcal U}_{l_2}$; we choose $d_{l_2}({\bf u}) = \frac{1}{2}\|{\bf u}  \|^2_2$ with a strong convexity parameter $\sigma=1$. Then $\forall g_i$, ${\bf u}^{l_2}_{g_i}$, which is a subvector of ${\bf u}^{l_2} $, can be calculated as ${\bf u}^{l_2}_{g_i}={\mathcal S}_2(\frac{\eta}{\mu}{\bf x}_{g_i})$ where ${\mathcal S}_2(\cdot)$ denotes the projection operator of projecting a vector $\bf a$ to a $\ell_2$ unit ball
\begin{align} 
{\mathcal S}_2(\bf a)=\left\{\begin{matrix}
\frac{\bf a}{\|{\bf a}\|_2},& \text{if  } {\|{\bf a}\|_2} > 1 \\ 
 {\bf a},&  \text{if  } {\|{\bf a}\|_2} \leq 1.
\end{matrix}\right.
\end{align}
Similarly, $\forall {\bf u} \in {\mathcal U}_{l_1},$ if we choose $d_{l_1}({\bf u}) = \frac{1}{2}\|{\bf u}  \|^2_2$, then ${\bf u}^{l_1}$ can be computed as ${\bf u}^{l_1}={\mathcal S}_1(\frac{\eta}{\mu}{\boldsymbol \nu})$ where ${\mathcal S}_1(\cdot)$ denotes the projection operator of projecting a vector $\bf a$ to an $\ell_{\infty}$ unit ball
\begin{align} 
{\mathcal S}_1({\bf a})=\left\{\begin{matrix}
1,& \text{if  } {a_i} > 1 , \forall i \\ 
 {a}_i,&  \text{if  } |{a_i}| \leq 1 , \forall i \\
 -1, & \text{if  } {a_i} <-1, , \forall i
\end{matrix}\right.
\end{align}
where $a_i$ is the $i$-th entry of ${\bf a}$.

Note that the dimension of ${\boldsymbol \nu}$ is a half of that for $\bf x$. Therefore, for the case of $\nabla h^{l_1}_{\mu}({\boldsymbol \nu})$, zero-padding is performed such that $\nabla h^{l_1}_{\mu}({\bf x}):=[\nabla h^{l_1}_{\mu}({\boldsymbol \nu})^T ,{\bf 0}^T]^T \in {\mathbb R}^{2N\times1}$, where ${\bf 0}$ is a ${\mathbb R}^{N\times1}$ zero vector, so that a new gradient $\nabla h^{l_1}_{\mu}({\bf x})$ can be used in the accelerated proximal gradient. This is acceptable only when parameter $r$ is taken small enough. Since ${\bf p} \ll {\bf s}$ holds in this case, the value of $\nu_{i}$ mainly comes from the contribution of $\bf s$, so that zero vector can be assigned as the partial derivative of $\bf p$.

\subsection{Accelerated Smoothing Proximal Gradient (ASPG)}
Now, we solve two "smoothed" versions of problem (\ref{glasso_variant})
\begin{align} \label{3smoothfuns}
\arg\min_{{\bf x} \in {\mathbb R}^n} \{ H_i({\bf x}) + \iota_{\mathcal X}({\bf x}) \},  i=1 \text{~or~} 2. 
\end{align} 
where $H_i({\bf x}):=f({\bf x})+h_{\mu}^{l_i}({\bf x})$, $i=1$ or $2$, and then its gradient is computed as $\nabla{H_i}({\bf x})=\nabla{f(\bf x)}+\eta {\bf u}^{l_i}$. 

Problem (\ref{3smoothfuns}) can be solved by the accelerated proximal gradient method  \cite{parikh2014proximal} in which a proximal operator is used:
 \begin{align} 
\text{prox}_{\iota}({\bf y})=\arg\min_{{\bf x} \in {\mathbb R}^n} \{ \frac{1}{2}\| {\bf y-x} \|^2 + \iota({\bf x})\}.  
\end{align}
In fact, the proximal operator $\text{prox}_{\iota_{\mathcal X}}(\bf y)$ of indicator function $ \iota_{\mathcal X}({\bf x})$ is the projection operator onto the set $\mathcal X$, $\Pi_{\mathcal X}(\bf x)$.
The ASPG method is summarized in the Algorithm \ref{alg::ASPG}.

\begin{algorithm}[htb]
  \caption{Accelerated Smoothing Proximal Gradient}
  \label{alg::ASPG}
  \begin{algorithmic}[1]
    \REQUIRE
      ${\bf x}^0={\bf x}^{1}={\bf 0}$; $\gamma=0.5$; $\mu=10^{-8}$; step-size $\alpha^0=1$;\\

    \ENSURE ($k\geq1$) Let $\alpha:=\alpha^{k-1}$. Compute \\
      ${\bf w}^{k+1}={\bf x}^k+\frac{k}{k+3}({\bf x}^k-{\bf x}^{k-1})$ 
    \REPEAT
      \STATE Compute $\nabla f({\bf w}^{k+1})={\bf G}^H({\bf G}{\bf w}^{k+1}-{\bf y})$,
      \STATE Compute $\nabla h_{\mu}^{l_i}({\bf w}^{k+1})=\eta {\bf u}^{l_2}$ if $ i=2$,
      \STATE Compute $\nabla h_{\mu}^{l_i}({\bf w}^{k+1})=\eta {\bf u}^{l_1}$ if $i=1$,
      \STATE  ${\bf z}=\Pi_{\mathcal X}({\bf w}^{k+1}-\alpha\nabla f({\bf w}^{k+1}) -\alpha \nabla h_{\mu}^{l_i}({\bf w}^{k+1}) )$,
      \STATE   Break if $F_i({\bf z}) \leq \hat{F}_i^{\alpha}({\bf z},{\bf w}^{k+1}) =F_i({\bf w}^{k+1})+(\nabla F_i({\bf w}^{k+1}) )^T({\bf z}-{\bf w}^{k+1}) +\frac{1}{2\alpha}\|{\bf z}-{\bf w}^{k+1}\|^2_2$,
      \STATE  Update 	 $\alpha:=\gamma \alpha$,
    \UNTIL $\alpha^k := \alpha$, ${\bf x}^{k+1}:= {\bf z}$
  \end{algorithmic} 
Note 1: ${\bf u}^{l_2}$ is composed of ${\bf u}^{l_2}_{g_i}={\mathcal S}_2(\frac{\eta}{\mu}{\bf w}^{k+1}_{g_i}), \forall g_i$.\\
Note 2: ${\bf u}^{l_1}=[{\mathcal S}_1(\frac{\eta}{\mu}{\boldsymbol \nu})^T,{\bf 0}^T]^T$ where $\nu_i=\| {\bf w}^{k+1}_{g_i} \|_2$, $\nu_i:$ $i$-th entry of ${\boldsymbol \nu}$
\end{algorithm}

\subsection{Convergence Analysis}

We show the convergence rate of the Algorithm 2 in the Lemma 2.
\begin{lemma}\label{conv_anal}
Suppose ${\bf x}^k$ is the $k$-th iterative solution in Algorithm \ref{alg::ASPG}, and $\bf x^*$ is the optimal solution of problem (\ref{glasso_variant}). Assume that $\epsilon$-approximation is required, i.e., $F({\bf x}^k)-F({\bf x}^*)\leq \epsilon$. If we set $\mu=\frac{\epsilon}{2D_{i}}$, where $D_i=\max_{{\bf u}\in {\mathcal U}_{l_i}}d_{l_i}({\bf u})$, then  
\begin{align}\label{epsilon_diff}
F({\bf x}^k)-F({\bf x}^*)\leq \frac{\epsilon}{2}+\frac{2(L_f+2\frac{D_i}{\epsilon\sigma})\|{\bf x}^0-{\bf x}^* \|^2}{(k+1)^2},
\end{align}
where $L_f$ is Lipschitz continuous gradient parameter of $f(\bf x)$. The number of iteration $k$ has an upper bound by 
\begin{align}\label{epsilon_appr}
\sqrt{\frac{4\| {\bf x}^0-{\bf x}^*  \|^2}{\epsilon}(L_f+\frac{2D_i}{\epsilon\sigma})}-1
\end{align} 
\end{lemma}

This lemma implies its convergence rate is ${\mathcal O}(\frac{1}{k})$. We cannot achieve convergence rate ${\mathcal O}(\frac{1}{k^2})$ of the accelerated proximal gradient method due to the smoothing process, but the convergence rate is better than that for subgradient methods with ${\mathcal O}(\frac{1}{\sqrt k})$ \cite{ben1989smoothing,shor2012minimization}.

\section{The EGT-based Primal-Dual Method}
 In ASPG, the smoothing paramter $\mu$ is chosen empirically and fixed. This leads to decrease the practical efficiency of ASPG. Thus, the excessive gap technique \cite{nesterov2005excessive} is employed to choose $\mu$ systematically in the framework of  primal-dual gradient symmetric formulations.

Let us consider the constrained optimization problem (\ref{glasso_std}) as follows:
\begin{equation} 
\begin{aligned}
& \underset{{\bf x}\in {\mathcal X}}{\text{arg min}}~~ F({\bf x})=\{ f({\bf x}) + h({\bf x}) \},  \\
& \text{s.t.}~~ {\mathcal X}=\{ {\bf x}=[{\bf s}^T,{\bf p}^T]^T : {\bf s}\geq0 , -r{\bf s}\leq {\bf p} \leq r{\bf s}\}.
\end{aligned}
\end{equation}
where $f({\bf x})=\frac{1}{2}||{\bf y}-{{\bf G}}{\bf x}||_2^2$, and $h({\bf x})= \eta||{\bf x}||_{2,1}=\max_{{\bf u}_2 \in {\mathcal U}_{l_2}}  \{ \eta  \langle{\bf x} , {\bf u}_2\rangle\}$. (Note that there are two reformulations of $h({\bf x})$ proposed in subsection III.A. The first one will be used for convenience to express the idea in this subsection.)

We know that $h({\bf x})$ is not strongly convex. 
And since ${\bf G}$ is a fat matrix, the error fitting function $f({\bf x})$ is not strongly convex either. Thus, we use  $f_r({\bf x})=||{\bf y}-{{\bf G}}{\bf x}||_2$ as a surrogate of $f({\bf x})$ such that it can be expressed in a $max$-structure form, and smoothed by using a strongly convex function, although $f_r({\bf x})$ is not differentiable everywhere. Thus, instead of solving (\ref{glasso_std}), we propose
\begin{equation} \label{glasso_std_ref}
\begin{aligned}
& \underset{{\bf x}\in {\mathcal X}}{\text{arg min}}~~ F({\bf x})=\{ f_r({\bf x}) + h({\bf x}) \},  \\
& \text{s.t.}~~ {\mathcal X}=\{ {\bf x}=[{\bf s}^T,{\bf p}^T]^T : {\bf s}\geq0 , -r{\bf s}\leq {\bf p} \leq r{\bf s}\}.
\end{aligned}
\end{equation}
Then, we will smooth not only the regularization term $h({\bf x})$, but also the new error fitting function $f_r({\bf x})$. This will lead to a closed form solution. Next, we will show how to achieve this goal by the excessive gap technique.

We can rewrite (\ref{glasso_std_ref}) into the following primal problem by using the dual norm definition:
\begin{align} \nonumber
 \arg & \min_{{\bf x}\in {\mathcal X}}  F({\bf x})= \\ 
& \{  \max_{{\bf u}=[{\bf u}_1^T,{\bf u}_2^T]^T,{\bf u}_1 \in {\mathcal U}_{2},{\bf u}_2 \in {\mathcal U}_{l_2}} \langle{\bf G}{\bf x} , {\bf u}_1\rangle  -\langle{\bf y} , {\bf u}_1\rangle +  \eta  \langle{\bf x} , {\bf u}_2\rangle \},
\end{align} \\
and its dual problem as
\begin{align} \nonumber
 \max_{{\bf u}=[{\bf u}_1^T,{\bf u}_2^T]^T,{\bf u}_1 \in {\mathcal U}_{2},{\bf u}_2 \in {\mathcal U}_{l_2}} & \Phi({\bf u}):= \\
& \{ -\langle{\bf y} , {\bf u}_1\rangle + \min_{{\bf x}\in {\mathcal X}} \langle{\bf G}{\bf x} , {\bf u}_1\rangle+\eta  \langle{\bf x} , {\bf u}_2\rangle\},
\end{align}
 where ${\bf u}$ is a dual variable vector composed of ${\bf u}_1$ and ${\bf u}_2$, which belong to ${\mathcal U}_{2}$, and ${\mathcal U}_{l_2}$, respectively, where 
\begin{align} \nonumber
{\mathcal U}_{2}=\{  {\bf u} \in {\mathbb R}^{M\times 1} : {\|{\bf u}\|_2\leq1}  \}.
 \end{align}
Since both $F({\bf x})$ and $\Phi({\bf u})$ are nondifferentiable, we can construct  a smoothing approximation of primal-dual problem as follows
 \begin{align} \label{sm_prim} \nonumber
 \min_{{\bf x}\in {\mathcal X}}&  F_{\mu_2}({\bf x}):= \\
& \{  \max_{{\bf u}=[{\bf u}_1^T,{\bf u}_2^T]^T}    \langle{\bf G}{\bf x} -{\bf y}, {\bf u}_1\rangle + \eta  \langle{\bf x} , {\bf u}_2\rangle - \frac{\mu_2}{2} \|{\bf u}\|^2_2 \},
\end{align}
\begin{align} \label{sm_dual} \nonumber 
 \max_{{\bf u}=[{\bf u}_1^T,{\bf u}_2^T]^T} & \Phi_{\mu_1}({\bf u}):= \\
& \{ -\langle{\bf y} , {\bf u}_1\rangle + \min_{{\bf x}\in {\mathcal X}} \langle{\bf G}{\bf x} , {\bf u}_1\rangle+\eta  \langle{\bf x} , {\bf u}_2\rangle + \frac{\mu_1}{2} \|{\bf x} \|^2_2  \} 
\end{align}
by using two strongly convex functions $d_1({\bf x})=\frac{1}{2} \|{\bf x} \|^2_2$, and $d_2({\bf u})=\frac{1}{2} \|{\bf u} \|^2_2$ with two smoothing parameters $\mu_1$, and $\mu_2$.\\
For the primal problem, denote ${\bf u}_{1,\mu_2}, {\bf u}_{2,\mu_2}$ as the unique optimal solution of  $F_{\mu_2}({\bf x})$, which can be derived in closed forms as
\begin{align}
&{\bf u}_{1,\mu_2}({\bf x})=Proj_{{\mathcal U}_{2}}(\frac{{\bf G}{\bf x} -{\bf y}}{\mu_2}), \\
 &{\bf u}_{2,\mu_2}({\bf x})=Proj_{{\mathcal U}_{l_2}}(\frac{\eta{\bf x}}{\mu_2}).
\end{align}
By Danskin's theorem \cite{bertsekas1999nonlinear}, the gradient of $F_{\mu_2}({\bf x})$ is computed as 
\begin{align}
  \nabla F_{\mu_2}({\bf x})= {\bf G}^H{\bf u}_{1,\mu_2}({\bf x})+\eta {\bf u}_{2,\mu_2}({\bf x})
\end{align}
with Lipschitz-continuous constant $L_1(F_{\mu_2}({\bf x}))=\frac{1}{{\mu_2}}\| [{\bf G}, \eta {\bf I} ]^H \|^2$.\\
Similarly, for the dual problem, denote ${\bf x}_{\mu_1}$ as the unique optimal solution of  $\Phi_{\mu_1}({\bf u})$, which can be derived in a closed form as
\begin{align}
{\bf x}_{\mu_1}({\bf u})=Proj_{{\mathcal X}}(-\frac{{\bf G}^H{\bf u}_1+\eta{\bf u}_2}{\mu_1}).
\end{align}
And the gradient of $\Phi_{\mu_1}({\bf u})$ is 
\begin{align}
\nabla \Phi_{\mu_1}({\bf u})= \begin{bmatrix}
     - {\bf y}  \\
      {\bf 0} 
    \end{bmatrix}+ \begin{bmatrix}
      {\bf G}{\bf x}_{\mu_1}(\bf u) \\
      \eta {\bf x}_{\mu_1}(\bf u) 
    \end{bmatrix}
\end{align}
 wth Lipschitz-continuous constant $L_2(\Phi_{\mu_1}({\bf u}))$ $=\frac{1}{{\mu_1}}\| [{\bf G}^H, \eta {\bf I} ]^H \|^2$ by Danskin's theorem. \\
Since we know that
\begin{itemize} 
\item $\Phi({\bf u}) \leq F({\bf x})$
\item By definition, $F_{\mu_2}({\bf x})\leq F({\bf x})$,  $\Phi({\bf u}) \leq \Phi_{\mu_1}({\bf u})$
\item Excessive gap condition (EGC) \cite{nesterov2005excessive} holds when, for certain ${\bf x}\in {\mathcal X}$ and ${{\bf u}=[{\bf u}_1^T,{\bf u}_2^T]^T,{\bf u}_1 \in {\mathcal U}_{2},{\bf u}_2 \in {\mathcal U}_{l_2}}$ with sufficiently large $\mu_1, \mu_2$, this inequality  occurs
\begin{align}\label{EGC}
F_{\mu_2}({\bf x})\leq \Phi_{\mu_1}({\bf u}).
\end{align}
\end{itemize}
Then, the following modified lemma can be derived:
\begin{lemma}
Let ${\bf x} \in {\mathcal X}$ and ${\bf u}=[{\bf u}_1^T,{\bf u}_2^T]^T,{\bf u}_1 \in {\mathcal U}_{2},{\bf u}_2 \in {\mathcal U}_{l_2}$ satisfy EGC. Then,
\begin{align} \nonumber
0 &\leq \max\{ F({\bf x}) - F^*, F^*- \Phi({\bf u}) \}\\  \nonumber
& \leq  \Phi({\bf u}) - F({\bf x}) \leq \mu_1 D_1+\mu_2 D_2+\mu_2 D_3
\end{align}
 where $D_1=\max_{{\bf x} \in {\mathcal X}} \| {\bf x} \|^2$, $D_2=\max_{{\bf u}_1 \in {\mathcal U}_{2}} \| {\bf u}_1 \|^2$, $D_3=\max_{{\bf u}_2 \in {\mathcal U}_{l_2}} \| {\bf u}_2 \|^2$.
\end{lemma}
By this modified lemma, EGC provides an upper bound of primal-dual pair $({\bf x, u})$ so that we can update iteratively the primal-dual pair $({\bf x, u})$ and keep satisfying EGC as $\mu_1, \mu_2$ approach to zero. We also apply the  primal  gradient mapping \cite{nesterov2005excessive}:
\begin{align} \nonumber
T_{\mu_2}({\bf x})=\arg\min_{{\bf z}\in {\mathcal X}} \{  \langle \nabla F_{\mu_2}({\bf x}), & {\bf z}-{\bf x}\rangle + \\
&\frac{1}{2} L_1(F_{\mu_2}({\bf x})) \| {\bf z}-{\bf x} \|^2 \}
\end{align} \\
and the dual gradient mapping:
\begin{align} \nonumber
T_{\mu_1}({\bf u})=\arg\min_{{\bf v}\in {\mathcal U}_2} \{  \langle \nabla \Phi_{\mu_1}({\bf x}),& {\bf v}-{\bf u}\rangle +\\ 
&\frac{1}{2} L_2(\Phi_{\mu_1}({\bf x})) \| {\bf v}-{\bf u} \|^2 \}
\end{align}
to choose some starting point when satisfying the EGC. In our case, they can be simplified in closed forms:
\begin{align} 
&T_{\mu_2}(\hat{\bf x})= Proj_{\mathcal X}({\bf x} - \frac{1}{L_1(F_{\mu_2})}\nabla F_{\mu_2}({\bf x})) \\
& T_{\mu_1}(\hat{\bf u})= Proj_{{\mathcal U}_2}({\bf u} - \frac{1}{L_2(F_{\mu_1})}\nabla \Phi_{\mu_1}({\bf u})).
\end{align}
By choosing feasibly initial points for primal and dual variables, the modified lemma for the primal part of iterative algorithms is proposed as follows:
\begin{lemma} 
 For a starting point ${\bf x}_0$, define 
\begin{align} \nonumber
\bar{\bf x}={T_{\mu_2}({\bf x}_0)} \in {\mathcal X}, \bar{\bf u}={\bf u}^*_{\mu_2}({\bf x}_0)=\begin{bmatrix}
     {\bf u}^*_{1,\mu_2}({\bf x}_0)  \\
     {\bf u}^*_{2,\mu_2}({\bf x}_0) 
    \end{bmatrix} \in {\mathcal U}:={\mathcal U}_2 \cup {\mathcal U}_{l_2},
\end{align}
 for an arbitrary $\mu_2 > 0$, and any $\mu_1 \geq L_1(F_{\mu_2})$. Fix $\tau \in(0,1)$ and choose $\mu_1^+ = (1-\tau)\mu_1$,
\begin{align} \nonumber
&\hat{\bf x} =   (1-\tau)\bar{\bf x} +\tau {\bf x}_{\mu_1}(\bar{\bf u}), \\ \nonumber
&\bar{\bf u}_+ = (1-\tau)\bar{\bf u}+ \tau {\bf u}_{\mu_2}(\hat{\bf x}),\\ \nonumber
&\bar{\bf x}_+ = T_{\mu_2}(\hat{\bf x})=Proj_{\mathcal X}(\hat{\bf x} - \frac{1}{L_1(F_{\mu_2})}\nabla F_{\mu_2}(\hat{\bf x})).
\end{align}
Then $(\bar{\bf x}_+ ,\bar{\bf u}_+ )$ satisfies EGC (\ref{EGC}) with smoothness parameter $\mu_1^+,\mu_2^+$ provided that $\tau$ is chosen by $\frac{\tau^2}{1-\tau}\leq\frac{\mu_1}{L_1(F_{\mu_2})}$.
\end{lemma} 

Thus, if EGC is satisfied for certain primal-dual pair, then the primal-dual pair can be updated iteratively when keeping satisfy the EGC as $\mu_1$ and $\mu_2$ go to zero. In other words, we can try to decrease $\mu_1$ with fixed $\mu_2$ for the primal problem; decrease $\mu_2$ with fixed $\mu_1$ for the dual problem.
The updates for primal-dual pair is summarized in the following Algorithm \ref{alg::EGT}. The convergence rate is of order ${\mathcal O}(\frac{1}{k})$ given in \cite{nesterov2005excessive}.
\renewcommand{\algorithmicrequire}{\textbf{Input:}}
    \renewcommand{\algorithmicensure}{\textbf{Step k:}}
    \renewcommand{\algorithmicuntil}{\textbf{return}}

\begin{algorithm}[htb]
  \caption{Excessive Gap Technique (EGT)-based Primal-Dual Method}
  \label{alg::EGT}
  \begin{algorithmic}[1]
    \REQUIRE
      $\mu_1$$=\| [{\bf G}^T, \eta {\bf I}]^T \| \sqrt{\frac{D_2+D_3}{D_1}}$, \\~~~
      $\mu_2=\| [{\bf G}^T, \eta {\bf I}]^T \| \sqrt{\frac{D1}{D_2+D_3}}$, \\~~~
      $\bar{\bf x}_0 =T_{\mu_2}({\bf x}_0), \bar{\bf u}_0={\bf u}_{\mu_2}({\bf x}_0)$

    \ENSURE ($k\geq 0$)   
      \STATE $\tau = \frac{2}{k+3}$ 	
      	\STATE If $k$: even, then  \\
 $\hat{\bf x} =   (1-\tau)\bar{\bf x} +\tau {\bf x}_{\mu_1}(\bar{\bf u})$\\
$\bar{\bf u}_+ = (1-\tau)\bar{\bf u}+ \tau {\bf u}_{\mu_2}(\hat{\bf x})$\\
$\bar{\bf x}_+ = T_{\mu_2}(\hat{\bf x})= Proj_{\mathcal X}(\hat{\bf x} - \frac{1}{L_1(F_{\mu_2})}\nabla F_{\mu_2}(\hat{\bf x}))$\\
 $ \mu_1^+ = (1-\tau)\mu_1, \mu_2^+ =\mu_2$	
      \STATE If $k$: odd, then  \\
 $\hat{\bf u} =   (1-\tau)\bar{\bf u} +\tau {\bf u}_{\mu_2}(\bar{\bf x})$\\
$\bar{\bf x}_+ = (1-\tau)\bar{\bf x}+ \tau {\bf x}_{\mu_1}(\hat{\bf u})$\\
$\bar{\bf u}_+ = T_{\mu_1}(\hat{\bf u})= Proj_{\mathcal U}(\hat{\bf u} - \frac{1}{L_2(F_{\mu_1})}\nabla \Phi_{\mu_1}(\hat{\bf u}))$ \\
$\mu_2^+ = (1-\tau)\mu_2, \mu_1^+ =\mu_1$	
  \end{algorithmic} 
\end{algorithm}

\section{Extension: Smoothed Dual Conic Formulation}

In the previous approaches for solving the constrained BPDN problem (\ref{glasso_std}), it is not natural to select a proper regularization parameter $\eta$. However, an estimate error $\epsilon$ for the error fitting term $f({\bf x})$ might be known based on SNRs. Thus, while only keeping the nonsmooth penalty function $h({\bf x})$ as an objective, formulating $f({\bf x})$ into a constraint is preferred. This leads to reformulating (\ref{glasso_std}) into a conical convex optimization problem.

\subsection{Primal-Dual Conic Formulations and the Smoothing}
 Instead of solving linear inequalities constrained BPDN problem (\ref{glasso_std})
\begin{align}
&\arg\min  f({\bf x})+ h({\bf x}) ,\\ \nonumber
&s.t.~~{{\bf x}\in {\mathcal X}}
\end{align}
where $f({\bf x}):=\frac{1}{2}||{\bf y}-{{\bf G}}{\bf x}||_2^2$, $h({\bf x}):= \eta||{\bf x}||_{2,1}$, ${\mathcal X}=\{ {\bf x}=[{\bf s}^T,{\bf p}^T]^T : {\bf s}\geq0 , -r{\bf s}\leq {\bf p} \leq r{\bf s}\}$.
 Inspired by \cite{becker2011templates},  a quadratically constrained with linear inequalities constraints problem is considered
\begin{align} 
& \arg\min_{{\bf x} } ||{\bf x}||_{2,1} \\ \nonumber
&s.t.~~ \| {\bf y}-{{\bf G}}{\bf x} \|_2 \leq \epsilon, \\ \nonumber
&~~~~~~{\bf x}\in {\mathcal X},
\end{align}
 since it is more natural to select an appropriate $\epsilon$ rather than an appropriate regularization parameter $\eta$. \\
Note that ${\mathcal X}$ is a set of elements satisfying linear inequalities, so it can be replaced by a matrix form representation $ {\bf C}{\bf x} \leq {\bf 0}$.
Then, let us consider the conic form of the primal problem 
\begin{align} \label{Prim_conic}
& \arg\min_{{\bf x} \in {\mathbb R}^{2N\times 1}} ||{\bf x}||_{2,1} \\ \nonumber
&s.t.~ ({\bf y}-{\bf G}{\bf x},\epsilon) \in {\mathcal K}^M_2 :=\{ ({\bf a}, b) \in {\mathbb C}^M \times {\mathbb R}: \|{\bf a} \|_2\leq b \}, \\  \nonumber
&~~~~~~ {\bf C}{\bf x} \leq {\bf 0},
\end{align}
and derive its dual by Lagrange multipliers
\begin{align}  
  \arg\max_{{\bf z} \in {\mathbb C}^{M\times 1},{\bf w}\geq {\bf 0}}  g({\bf z},{\bf w}), \\ \nonumber
\end{align}
where $g({\bf z},{\bf w})=\inf_{\bf x} ||{\bf x}||_{2,1} -  \langle {\bf z}, {\bf y}-{\bf G}{\bf x} \rangle - \epsilon \| {\bf z}\|_2 +   \langle {\bf w}, {\bf C}{\bf x} \rangle$. \\
Note that both objectives are nonsmooth in the primal and dual formulation. So, we smooth $||{\bf x}||_{2,1}$ by adding the strongly convex \textit{prox-function} $d({\bf x})=\frac{\sigma \mu}{2}\| {\bf x} -{\bf x}_0 \|^2_2$ with a smoothing parameter $\mu$ and a strong convexity parameter $\sigma =1$. ${\bf x}_0$ is denoted as the \textit{prox-center} of $d({\bf x})$
\begin{align}  
  {\bf x}_0 =  \arg\min_{{\bf x} \in {\mathcal X}} d({\bf x}).  \nonumber
\end{align}
In this way, the smoothed dual problem is given by
\begin{align}  
  \arg\max_{{\bf z} \in {\mathbb C}^{M\times 1},{\bf w}\geq {\bf 0}}  g_{\mu}({\bf z},{\bf w}), \nonumber
\end{align}
where 
\begin{align}  \nonumber
 g_{\mu}({\bf z},{\bf w})=\inf_{\bf x} ||{\bf x}||_{2,1} & +\frac{\mu}{2}\| {\bf x} -{\bf x}_0 \|^2_2 \\ 
&-  \langle {\bf z}, {\bf y}-{\bf G}{\bf x} \rangle - \epsilon \| {\bf z}\|_2 +   \langle {\bf w}, {\bf C}{\bf x} \rangle   \nonumber
\end{align}  
  is a smooth function over $\bf x$.
The optimal solution of $g_{\mu}({\bf z},{\bf w})$ is unique because of the strong convexity of $d({\bf x})$.
Define ${\bf x}({\bf z},{\bf w})$ as the optimal solution of $g_{\mu}({\bf z},{\bf w})$ which is computed as
\begin{align}  
  {\bf x}({\bf z},{\bf w}) = GST( {\bf x}_0 - \frac{1}{\mu} ({\bf G}^H{\bf z}+{\bf C}^H{\bf w}), \frac{1}{\mu} ), \nonumber
\end{align}  
where a group-soft-thresholding operator $ GST({\bf x},t)$ of ${\bf x}=[{\bf s}^T, {\bf p}^T]^T \in {\mathbb R}^{2N}$ is defined as
\begin{align}  \nonumber
& GST({\bf x},t) \overset{\Delta}{=} \\
&  \frac{[x_i, x_{i+N}]}{\sqrt{x_i^2+x_{i+N}^2}}  \max\{  \sqrt{x_i^2+x_{i+N}^2}- t, 0 \}, 1\leq i \leq N  .
\end{align}
We rewrite the smoothed dual problem as
\begin{align}  \label{sm_dual_p}
  \arg\min_{{\bf z} \in {\mathbb C}^{M\times 1},{\bf w}\geq {\bf 0}} - \bar{g}_{\mu}({\bf z},{\bf w}) ={g}_{sm}({\bf z},{\bf w}) +h({\bf z}),   
\end{align}
where 
\begin{align}\nonumber
 g_{sm}({\bf z},{\bf w})= & -||{\bf x}({\bf z},{\bf w})||_{2,1} -\frac{\mu}{2}\| {\bf x}({\bf z},{\bf w}) -{\bf x}_0 \|^2_2 \\  \nonumber
&+  \langle {\bf z}, {\bf y}-{\bf G}{\bf x}({\bf z},{\bf w}) \rangle -   \langle {\bf w}, {\bf C}{\bf x}({\bf z},{\bf w}) \rangle , \\ \nonumber
 h({\bf z})=\epsilon \| {\bf z} \|_2.
\end{align}

\subsection{Smoothed Dual Conic Optimization (SDCO) Solver}
The problem (\ref{sm_dual_p}) we try to solve is in a composite form with smooth part $g_{sm}$ and nonsmooth part $h$.
The smoothed part $ g_{sm}({\bf z},{\bf w})$  is differentiable and its gradient is computed as  $\nabla g_{sm}({\bf z},{\bf w})=\begin{bmatrix}
  {\bf y}- {\bf G}{\bf x}({\bf z},{\bf w}) \\
  - {\bf C}{\bf x}({\bf z},{\bf w})
\end{bmatrix}$
in accordance with Danskin's theorem.

Then, the generalized gradient projection method \cite{auslender2006interior,Paul2008on} is applied to solve (\ref{sm_dual_p}) by updating
\begin{align}  \label{ggpm} \nonumber
 &({\bf z}_{k+1},{\bf w}_{k+1}) = \arg\min_{{\bf z} \in {\mathbb C}^{M\times 1},{\bf w}\geq {\bf 0}}  {g}_{sm}({\bf z}_k,{\bf w}_k) + \\  \nonumber
 &\langle \nabla  {g}_{sm}({\bf z}_k,{\bf w}_k), ({\bf z}-{\bf z}_k,{\bf w}-{\bf w}_k ) \rangle 
+ \frac{L_k}{2}\| ({\bf z}-{\bf z}_k,{\bf w}-{\bf w}_k ) \|^2 \\  
&+\epsilon \| {\bf z} \|_2 ,
\end{align}
where $L_k$ is the inverse of step size $t_k$. References \cite{nesterov2013introductory,nemirovskii1983problem} show that $\epsilon$-optimality can be achieved in ${\mathcal O}(1/\epsilon)$ iterations if $t_k$ is selected properly.
Actually, a closed form solution for $({\bf z}_{k+1},{\bf w}_{k+1}) $ can be derived as 
\begin{align}  \nonumber
  &{\bf z}_{k+1}\\ 
 &=\arg\min_{\bf z}~ \langle {\bf y}- {\bf G}{\bf x}({\bf z},{\bf w}),  {\bf z} - {\bf z}_k \rangle +  \frac{L_k}{2}\| {\bf z}-{\bf z}_k \|^2  +\epsilon \| {\bf z} \|_2 \\  \nonumber
 &= Shrink( {\bf z}_k - \frac{1}{L_k} ({\bf y}- {\bf G}{\bf x}({\bf z},{\bf w})), \frac{2 \epsilon}{L_k} ),  \nonumber
\end{align}  
\begin{align}  
  {\bf w}_{k+1} &=\arg\min_{{\bf w}\geq0} ~ \frac{2}{L_k}  \langle -{\bf C}{\bf x}({\bf z},{\bf w}),  {\bf w} - {\bf w}_k \rangle +  \frac{L_k}{2}\| {\bf w}-{\bf w}_k \|^2   \\ 
 &= {\bf w}_k + \frac{1}{L_k}{\bf C}{\bf x}({\bf z},{\bf w}), \nonumber
\end{align}  
where an $l_2$-shrinkage operation $Shrink({\bf x},t)$ is defined as
\begin{align}  \nonumber
 Shrink({\bf x},t) & \overset{\Delta}{=} \max\{ 1 - \frac{t}{\| {\bf x} \|_2}, 0 \} \cdot {\bf x}  \\
& = \left\{
\begin{array}{rcl}
0,      &      & {     \| {\bf x} \|_2   \leq t  }  \\
(1-t/{\| {\bf x} \|_2}) \cdot {\bf x},    &      & {  \| {\bf x} \|_2  > t  }\\
\end{array} \right. .
\end{align}
The right-hand side is first-order approximation of (\ref{ggpm}), and satisfies an upper bound property
\begin{equation}\label{upp_bdd_L}
\begin{aligned} 
- \bar{g}_{\mu}({\bf z}_{k+1}, &{\bf w}_{k+1} ) \leq \\
& \langle \nabla  {g}_{sm}({\bf z}_k,{\bf w}_k), ({\bf z}_{k+1}-{\bf z}_k,{\bf w}_{k+1}-{\bf w}_k ) \rangle  \\  
&+  \frac{L_k}{2}\| ({\bf z}_{k+1}-{\bf z}_k,{\bf w}_{k+1}-{\bf w}_k ) \|^2 +\epsilon \| {\bf z}_{k+1} \|_2,
\end{aligned} 
\end{equation}
which holds for sufficiently large $L_k$. Typically, if $L_k\geq L, \forall k$, then the upper bound (\ref{upp_bdd_L}) holds, where $L$ is Lipschitz constant. Under those assumptions, $\epsilon$-optimality can be achieved in ${\mathcal O}(L/\epsilon)$ iterations by performing (\ref{ggpm}). A variation of the generalized gradient projection method proposed by Nesterov, which is an optimal first-order method with ${\mathcal O}(L/\sqrt{\epsilon})$ iterations, is used instead of (\ref{ggpm}). The approach is summarized in Algorithm \ref{alg::SDCO}.
\renewcommand{\algorithmicrequire}{\textbf{Input:}}
    \renewcommand{\algorithmicensure}{\textbf{Step k:}}
    \renewcommand{\algorithmicuntil}{\textbf{return}}

\begin{algorithm}[htb]
  \caption{Smoothed Dual Conic Optimization}
  \label{alg::SDCO}
  \begin{algorithmic}[1]
    \REQUIRE
      ${\bf x}_0={\bf 0}$, ${\bf z}_0={\bf 0}$, ${\bf w}_0={\bf 0}$, \\
~~~~$\mu=1$, $L_k=1$, $c_0=1$,$\gamma=0.5$ \\
~~~~${\bf s}_0=[{\bf z}_0^T, {\bf w}_0^T]^T$

    \ENSURE ($k\geq 0$) Let $ L:=L_k$,  
      \STATE $[{\bf z}_k^T, {\bf w}_k^T]^T=(1-c_k){\bf s}_k+c_k[{\bf z}_k^T, {\bf w}_k^T]^T$ 	
      \STATE ${\bf x}({\bf z}_k,{\bf w}_k)=\inf_{\bf x} ||{\bf x}||_{2,1} +\frac{\mu}{2}\| {\bf x} -{\bf x}_0 \|^2_2 -  \langle {\bf z}_k, {\bf y}-{\bf G}{\bf x} \rangle - \epsilon \| {\bf z}\|_2 +   \langle {\bf w}_k, {\bf C}{\bf x} \rangle$
       
       \REPEAT
      \STATE 
 $({\bf z}_{k+1},{\bf w}_{k+1}) = \arg\min_{{\bf z} ,{\bf w}\geq {\bf 0}}  
 \langle \nabla  {g}_{sm}({\bf z}_k,{\bf w}_k), ({\bf z}-{\bf z}_k,{\bf w}-{\bf w}_k ) \rangle 
+ \frac{L}{2}\| ({\bf z}-{\bf z}_k,{\bf w}-{\bf w}_k ) \|^2  
+\epsilon \| {\bf z} \|_2 $	 
      \STATE   Break if $g_{sm}({\bf z}_{k+1},{\bf w}_{k+1}) \leq  g_{sm}({\bf z}_{k},{\bf w}_{k})+$ \\ $\langle \nabla  {g}_{sm}({\bf z}_k,{\bf w}_k), ({\bf z}-{\bf z}_k,{\bf w}-{\bf w}_k ) \rangle + \frac{L}{2}\| ({\bf z}-{\bf z}_k,{\bf w}-{\bf w}_k ) \|^2   $,
      \STATE  Update 	 $L:=L/\gamma $,
    \UNTIL $L_k := L$

       \STATE ${\bf s}_k=(1-c_k){\bf s}_k+c_k[{\bf z}_{k+1}^T, {\bf w}_{k+1}^T]^T $, $c_{k+1}=\frac{2}{1+\sqrt{1+4/c_k^2}}$
           	
  \end{algorithmic} 
\end{algorithm}
It is noted that the smaller smoothing parameter $\mu$, the better is the accuracy performance.
On the other hand, the continuation scheme, which was proposed in NESTA \cite{becker2011nesta}, improves the convergence rate. Accordingly, a sequence of subproblems  is solved by Algorithm \ref{alg::SDCO} with decreasing smoothing parameters $\mu_k$. Each result of subproblems feeds into the next round. The standard continuation scheme combined with Algorithm \ref{alg::SDCO} is listed below:
\renewcommand{\algorithmicrequire}{\textbf{Input:}}
    \renewcommand{\algorithmicensure}{\textbf{Step j:}}
    \renewcommand{\algorithmicuntil}{\textbf{return}}

\begin{algorithm}[htb]
\caption{Standard Continuation}
  \begin{algorithmic}[1]
    \REQUIRE
      ${\mathit X}_0$: the set of variables in Algorithm \ref{alg::SDCO}, \\
~~~~~$\mu_0 =1$, $\alpha=0.5$  

    \ENSURE ($j\geq 0$)   
      \STATE ${\mathit X}_{j+1} \leftarrow \textit{Algorithm \ref{alg::SDCO}}$	
      \STATE $\mu_{j+1}=\alpha \mu_j$
      	
  \end{algorithmic} 
\end{algorithm}

\section{Numerical Results}
In this section, the off-grid DoA estimation is conducted to demonstrate the performance of the proposed methods. The two proposed accelerated smoothing proximal gradient methods are designated as ASPG-L2 (using $ h^{l_2}_{\mu}({\bf x})$) and ASPG-L1 (using $\nabla h^{l_1}_{\mu}({\boldsymbol \nu})$), the consensus ADMM method is designated as C-ADMM. the variant of excess-gap technique method is called EGT-based, and the variant of smoothed dual conic optimization method with continuation is called SDCO-Ct. We also solve problem (\ref{glasso_std}) by using CVX packages. The CVX method implemented by the interior point method can be viewed as a benchmark, which is used to evaluate the estimation performance degradation caused by smoothing in the proposed methods. The estimation errors of these  methods are compared with the same for the MUSIC estimator, M+LFBF and the CRLB. Consider $K=2$ uncorrelated source signals from DoAs ${\boldsymbol \theta}=[13.2220, 28.6022]$ degree impinging on a uniform linear array of $M=8$ sensors with half-wavelength interelement spacing.  The two sources are randomly generated with normal distribution of zero mean and variance $\sigma_s^2$. The noise term is i.i.d. AWGN with zero mean and variance $\sigma_n^2$. We use one hundred snapshots to estimate the covariance matrix. The size $N$ of search grid is set to 360 with $r=0.25$ degree, which is used for all methods. One hundred realizations are performed at each SNR.
In the ASPG method, the decreasing factor is $\gamma=0.5$, and smoothing parameter is chosen as $\mu=10^{-8}$. In the EGT-based method, the two smoothing parameters $\mu_1, \mu_2$ are controlled by $\tau=\frac{2}{k+3}$, where $k$ is the iteration number. In the SDCO-Ct method, the initial value of smoothing parameter is set to one, and sequentially reduced by multiplying with $0.5$ at each step in the outer loop. All the other parameter settings can be referred in the Algorithm blocks. 

\begin{figure} 
\begin{center}
\includegraphics[width=\columnwidth]{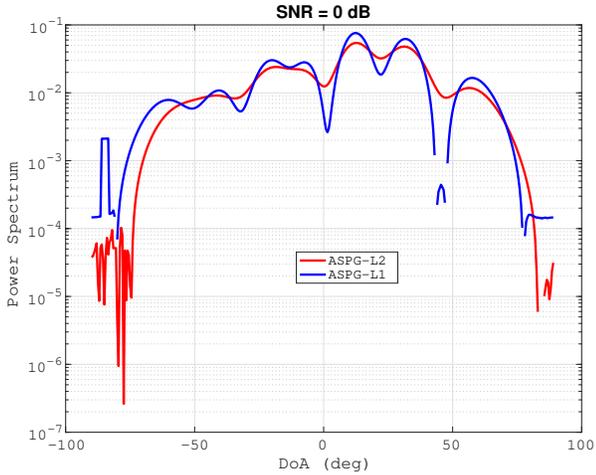}
\caption{Power Spectrum versus DoA for ASPG-L2, and ASPG-L1 at SNR = 0 dB.}\label{fig:ch3 estimation exp}
\end{center}
\end{figure}

\begin{figure} 
\begin{center}
\includegraphics[width=\columnwidth]{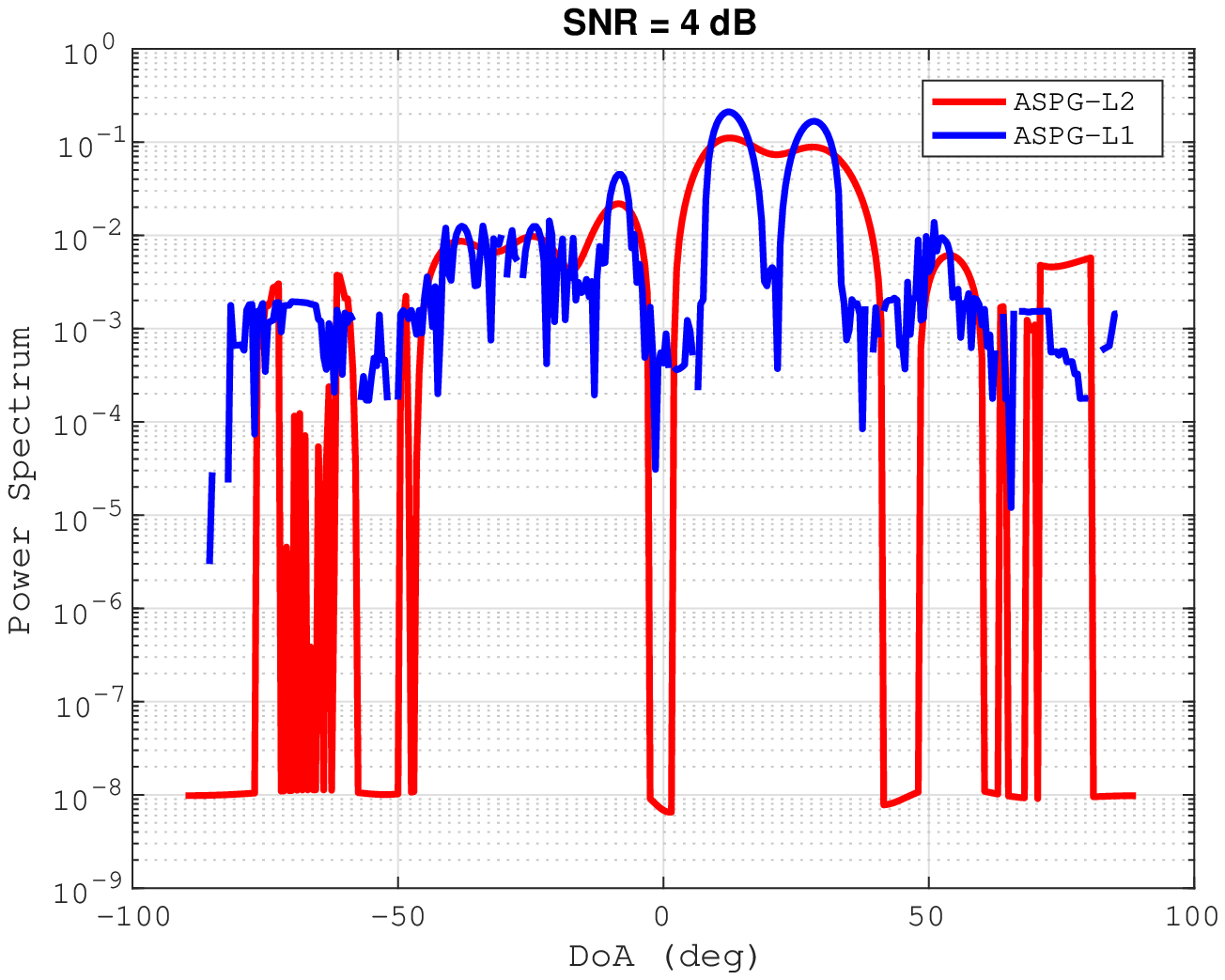}
\caption{Power Spectrum versus DoA for ASPG-L2, and ASPG-L1 at SNR = 4 dB.}\label{fig:ch3 estimation exp_snr4}
\end{center}
\end{figure}

\begin{figure}[htb]
\begin{center}
\includegraphics[width=\columnwidth]{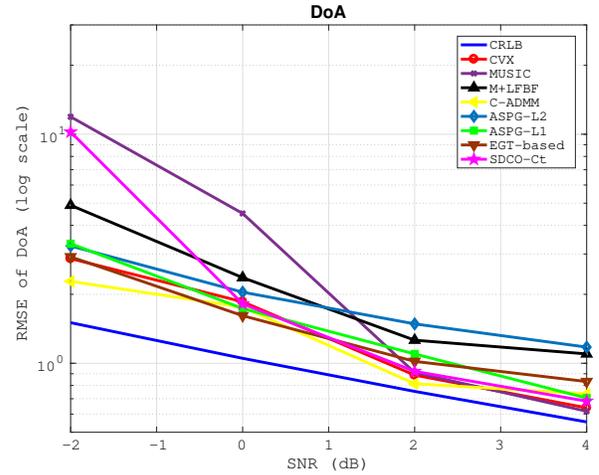}
\caption{RMSE of DoA estimation versus SNR.}\label{fig:ch3 RMSE of DoA}
\end{center}
\end{figure}

\subsection{DoA Resolution of Two Reformulated Group-sparsity Penalties}
The resolution ability of two reformulated group-sparsity penalties (using $ h^{l_2}_{\mu}({\bf x})$, and $\nabla h^{l_1}_{\mu}({\boldsymbol \nu})$)  is verified with the ASPG method.
In Figure \ref{fig:ch3 estimation exp}, the estimated power spectrum of ASPG methods is presented at SNR $=0$ dB. Due to the smoothing process, both have lost their sparsity. However, the two peaks of ASPG-L1 are more separated than ASPG-L2. In other words, ASPG-L1 estimator owns higher DoA resolution. In Figure \ref{fig:ch3 estimation exp_snr4}, at SNR $=4$ dB, the resolution ability of ASPG-L1 estimator gets improved compared with the case of SNR $=0$ dB, while ASPG-L2 estimator does not. As can be seen in Figure \ref{fig:ch3 estimation exp_snr4}, the shape of two major peaks of ASPG-L1 is sharper, and much more separated.

\subsection{Accuracy of Off-Grid DoA Estimation}
The accuracy performance of off-grid DoA estimation for the proposed methods is presented by the root-mean-square-error (RMSE) of DoA estimation, which is defined as $(E[\frac{1}{K}\| \hat{{\boldsymbol \theta}} - {\boldsymbol \theta} \|_2^2])^{\frac{1}{2}}$. 
Noted that since we show the DoA resolution of the second reformulation ($\nabla h^{l_1}_{\mu}({\boldsymbol \nu})$) is better than the first one, we perform the EGT-based method by adopting the second reformulated group-sparsity penalty in order to get better performance.
As seen in Figure \ref{fig:ch3 RMSE of DoA}, the RMSE of CVX, C-ADMM, ASPG-L1, ASPG-L2,  EGT-based, SDCO-Ct are almost the same and better than MUSIC at SNR $=0$ dB. When SNRs are low, the performance degradation mainly comes from the bad estimation of nonzero term locations in the sparse vector ${\bf x}=[{\bf s}^T,{\bf p}^T]^T\in {\mathbb R^{2N\times 1}}$, where ${\bf p}={\boldsymbol \beta}\odot {\bf s}$. We notice that the RMSE of SDCO-Ct, and M+LFBF get worse at SNR $=-2$ dB, which also indicates that their resolution ability becomes weaker. \\
When SNRs are high, if the RMSE performance cannot approach CRLB, this means that the estimation of the off-grid DoA vector ${\boldsymbol{\beta}}$ is not satisfied.
At SNR $=2$, and 4 dB, the performance of ASPG-L1, CVX, ADMM, EGT-based, SDCO-Ct, and MUSIC is better than ASPG-L2, and M+LFBF. The reason of bad performance in the ASPG-L2 is that the sparse property of group-sparsity penalty $\| {\bf x}_{g_i} \|_2$ is lost during the smoothing process by only using the property that the dual norm of $\ell_2$ norm is also $\ell_2$ norm so that sparsity is not promoted in this way. Thus, a satisfying estimation of ${\boldsymbol \beta}$ cannot be obtained.

\begin{figure*}[htp]
  \centering
  \subfigure[MUSIC]{\includegraphics[width=\columnwidth,height=13.5em]{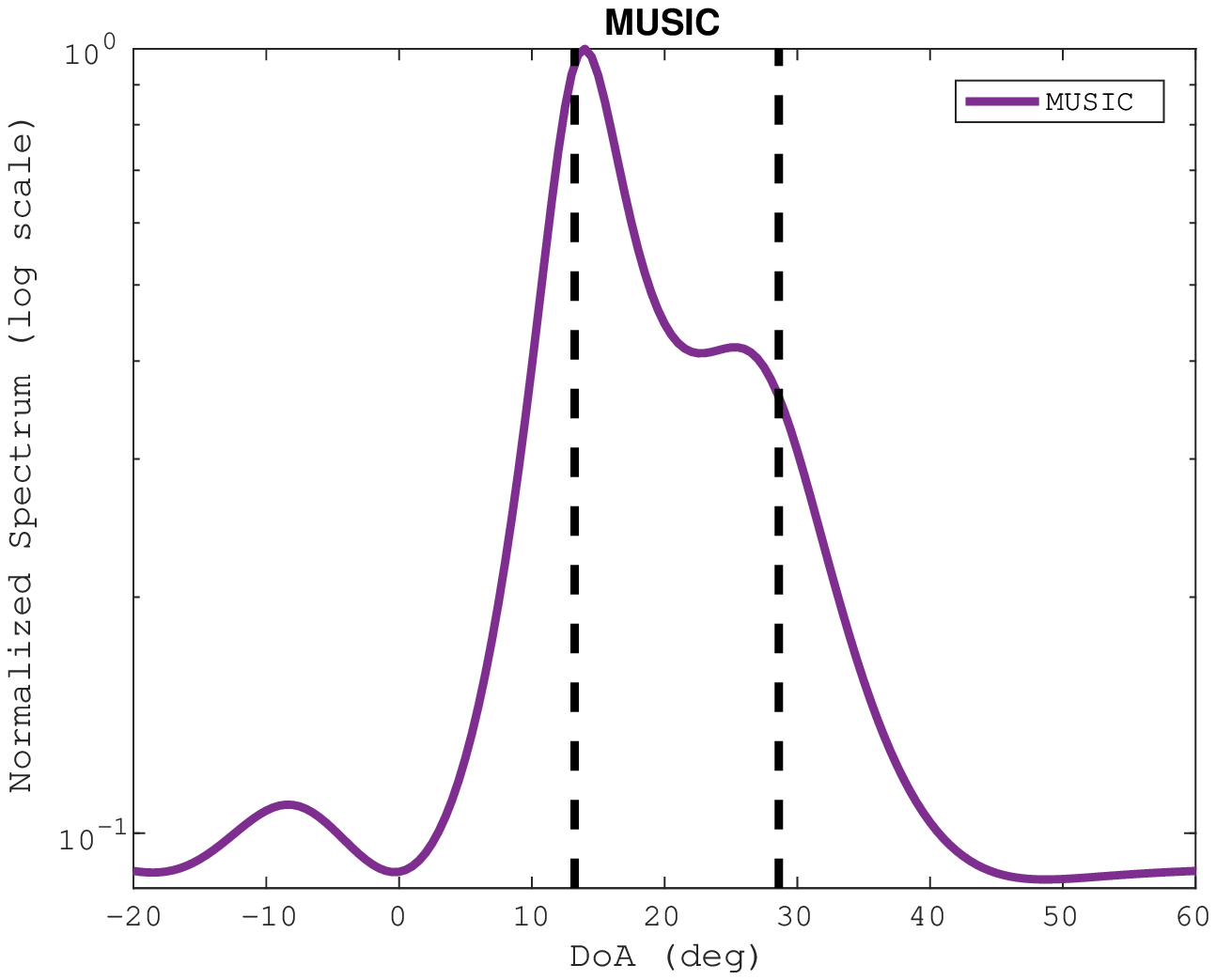}}\hfill
  \subfigure[M+LFBF]{\includegraphics[width=\columnwidth,height=13.5em]{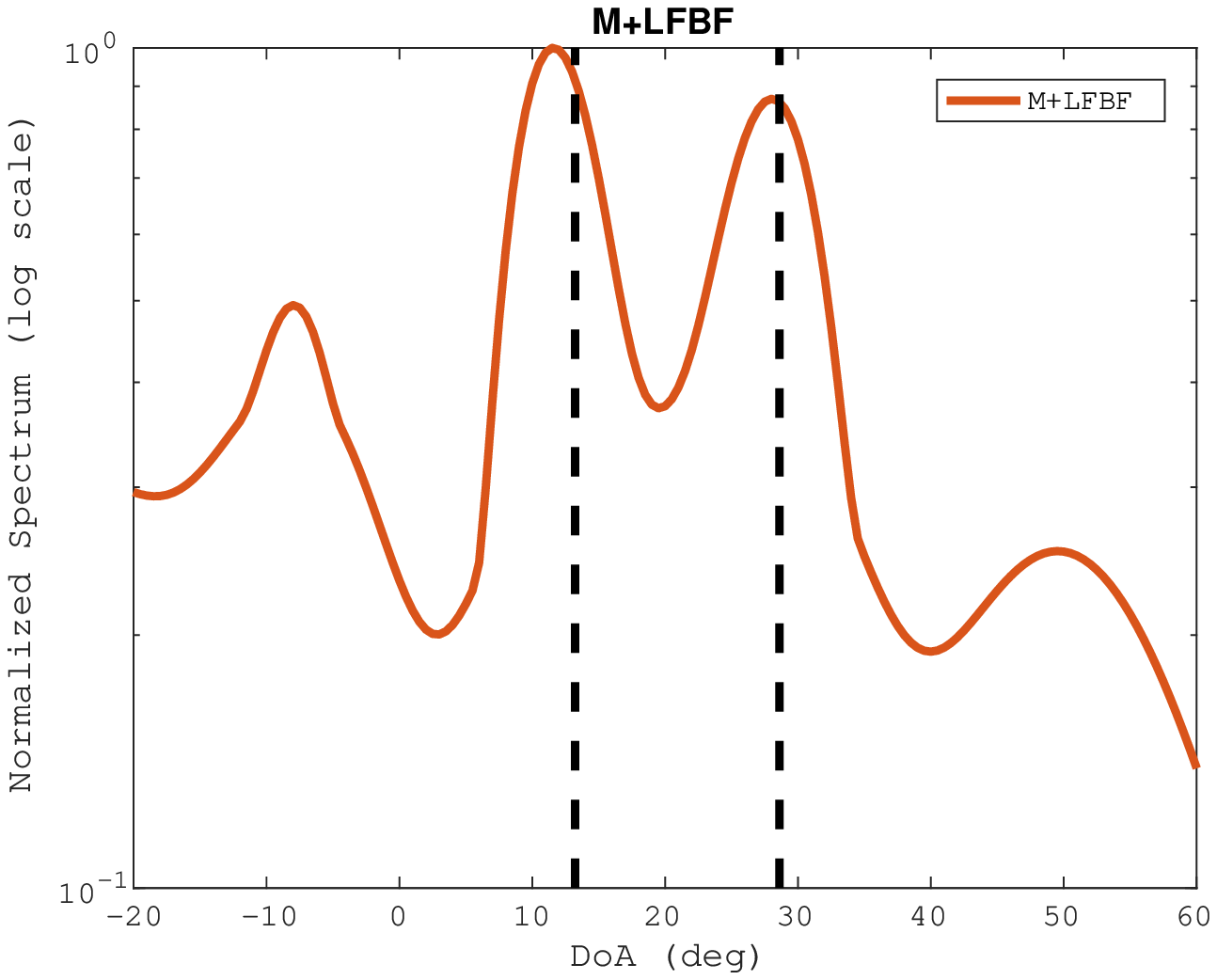}}\hfill
  \subfigure[ASPG-L1]{\includegraphics[width=\columnwidth,height=13.5em]{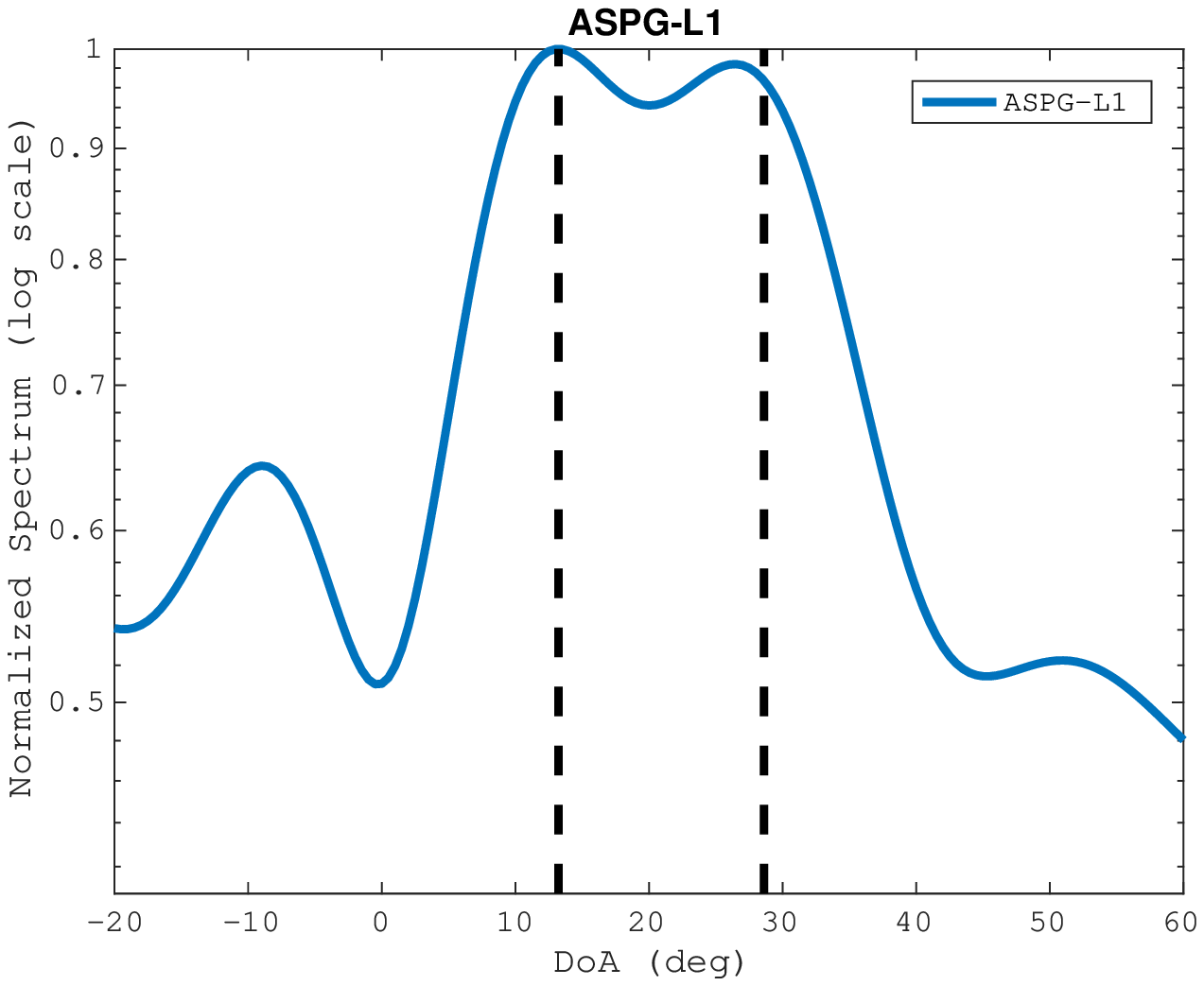}}\hfill
  \subfigure[EGT-based]{\includegraphics[width=\columnwidth,height=13.5em]{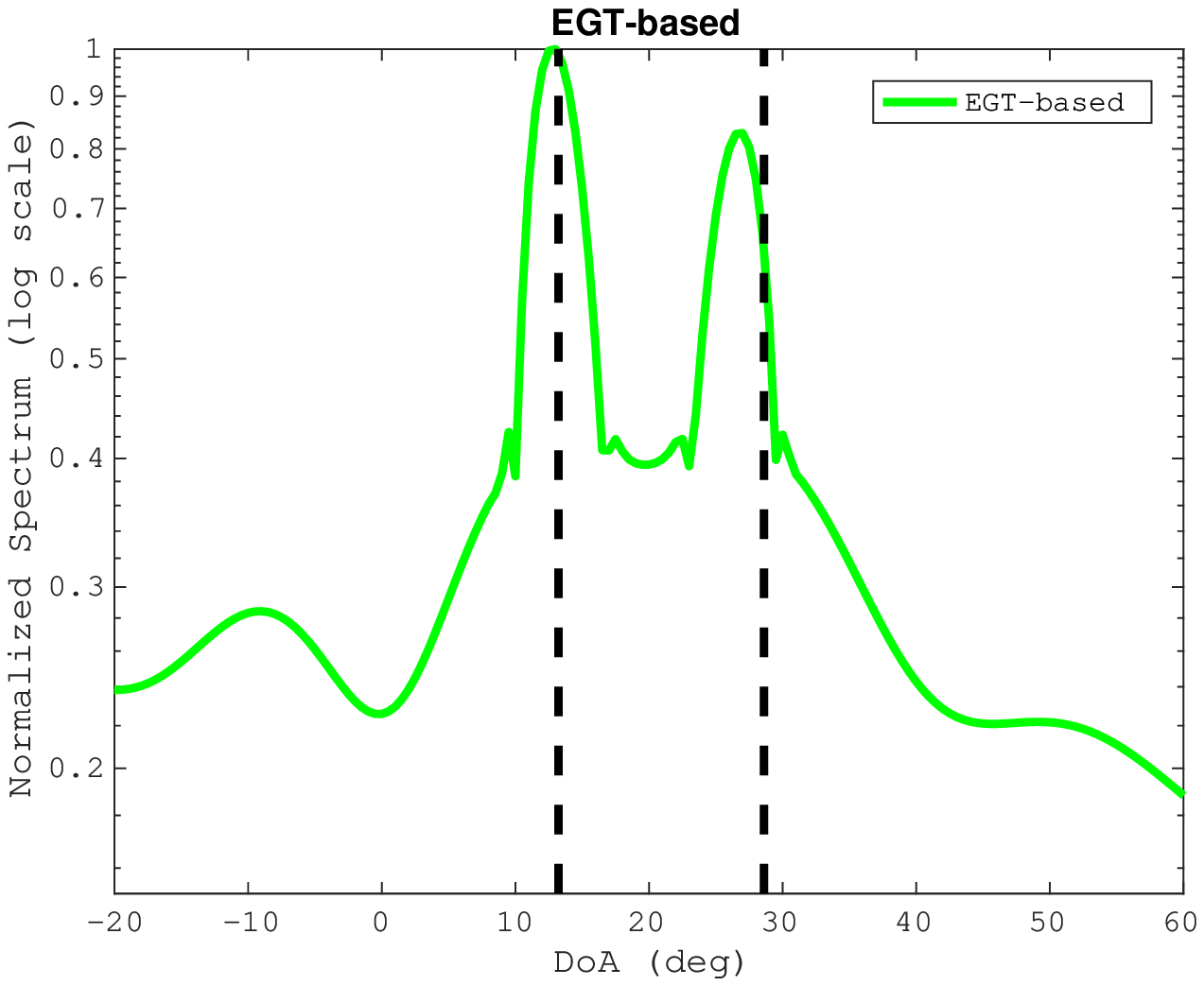}}\hfill
  \subfigure[SDCO-Ct]{\includegraphics[width=\columnwidth,height=13.5em]{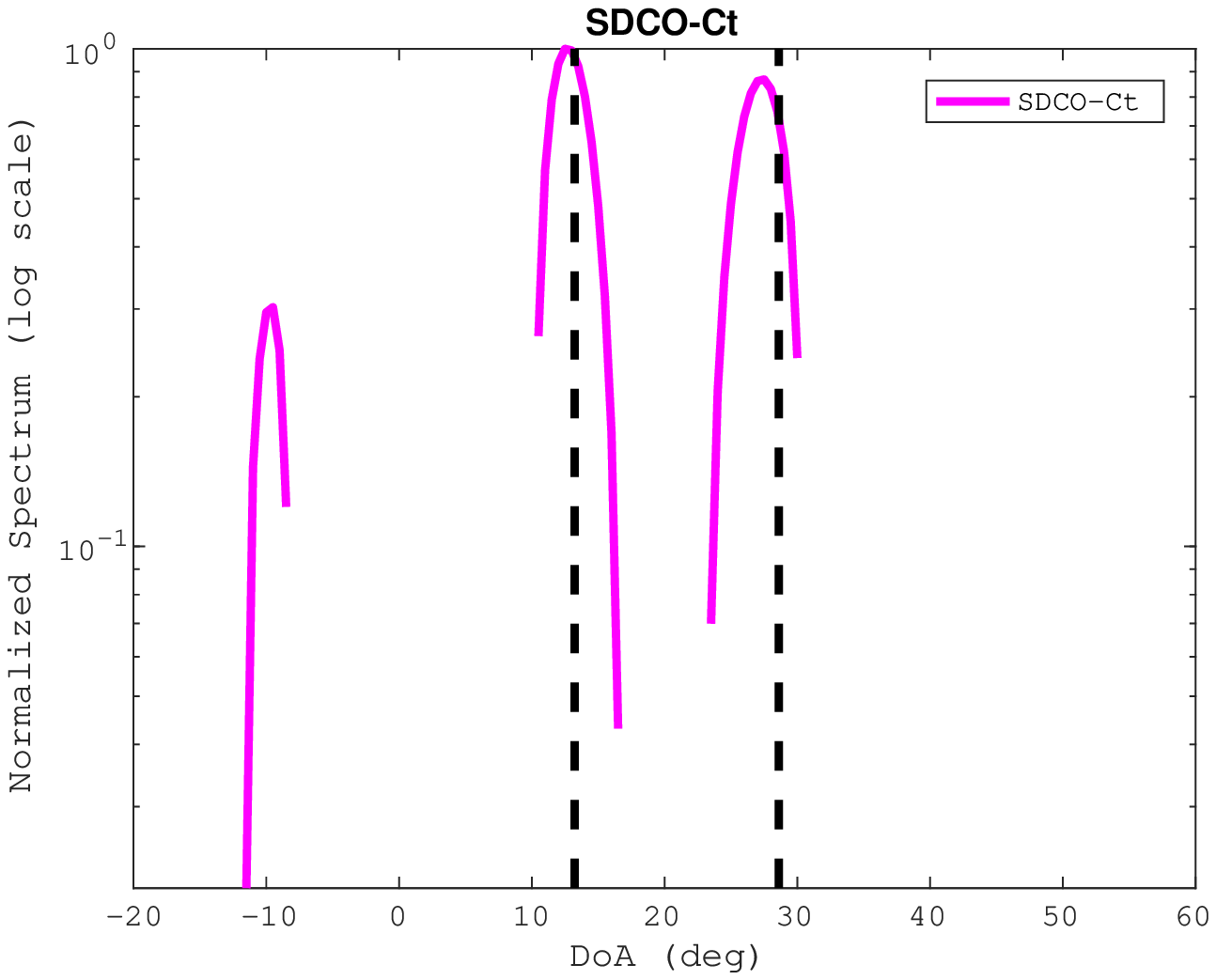}}\hfill
  \subfigure[C-ADMM]{\includegraphics[width=\columnwidth,height=13.5em]{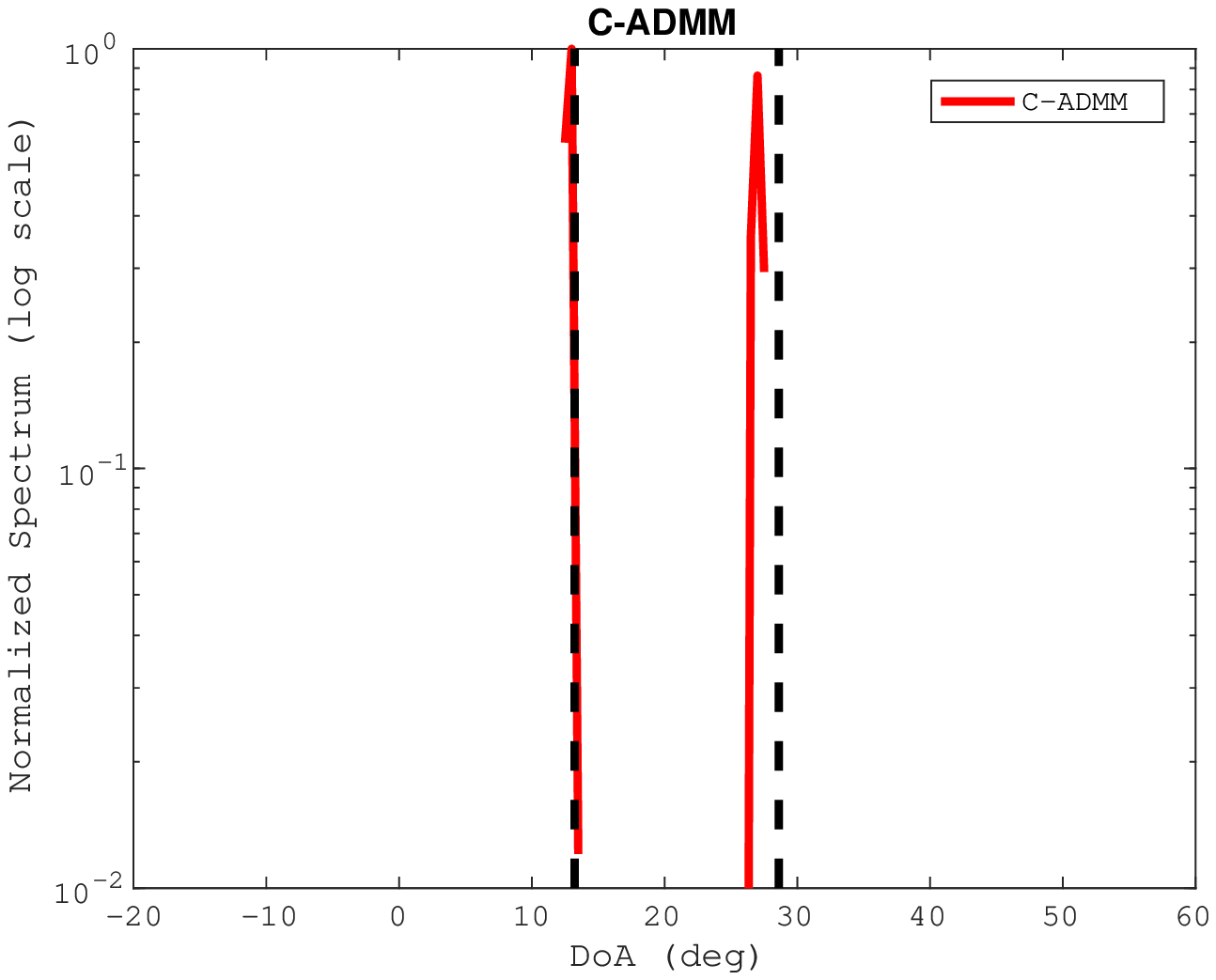}}\hfill
\caption{DoA resolution performance at SNR = 0 dB, $M=8$. These methods are MUSIC, M+LFBF, ASPG-L1, EGT-based, SDCO-Ct, and C-ADMM.} \label{fig:compare}
\end{figure*}

\subsection{DoA Resolution Performance}

In this numerical experiment, the resolution test is performed to demonstrate the ability of detecting two closely located DoAs for the proposed methods at SNR $=0$ dB by checking the normalized spectra. In Figure \ref{fig:compare}, the DoA resolution of MUSIC is worse than all the others because it almost cannot detect the second DoA. Due to the smoothing process, ASPG-L1, EGT-based, and SDCO-Ct lose the sparse property of group-sparsity penalty so that the shape of two major detected peaks is not sharp as C-ADMM. However, instead of using fixed smoothing parameters in the ASPG method, the EGT-based, and SDCO-Ct method use different approaches to sequentially reduce the smoothing parametersso that the resolution ability is improved. The sharpness of two peaks of SDCO-Ct is closer to C-ADMM compared with all the others.

\begin{figure}
\begin{center}
\includegraphics[width=\columnwidth]{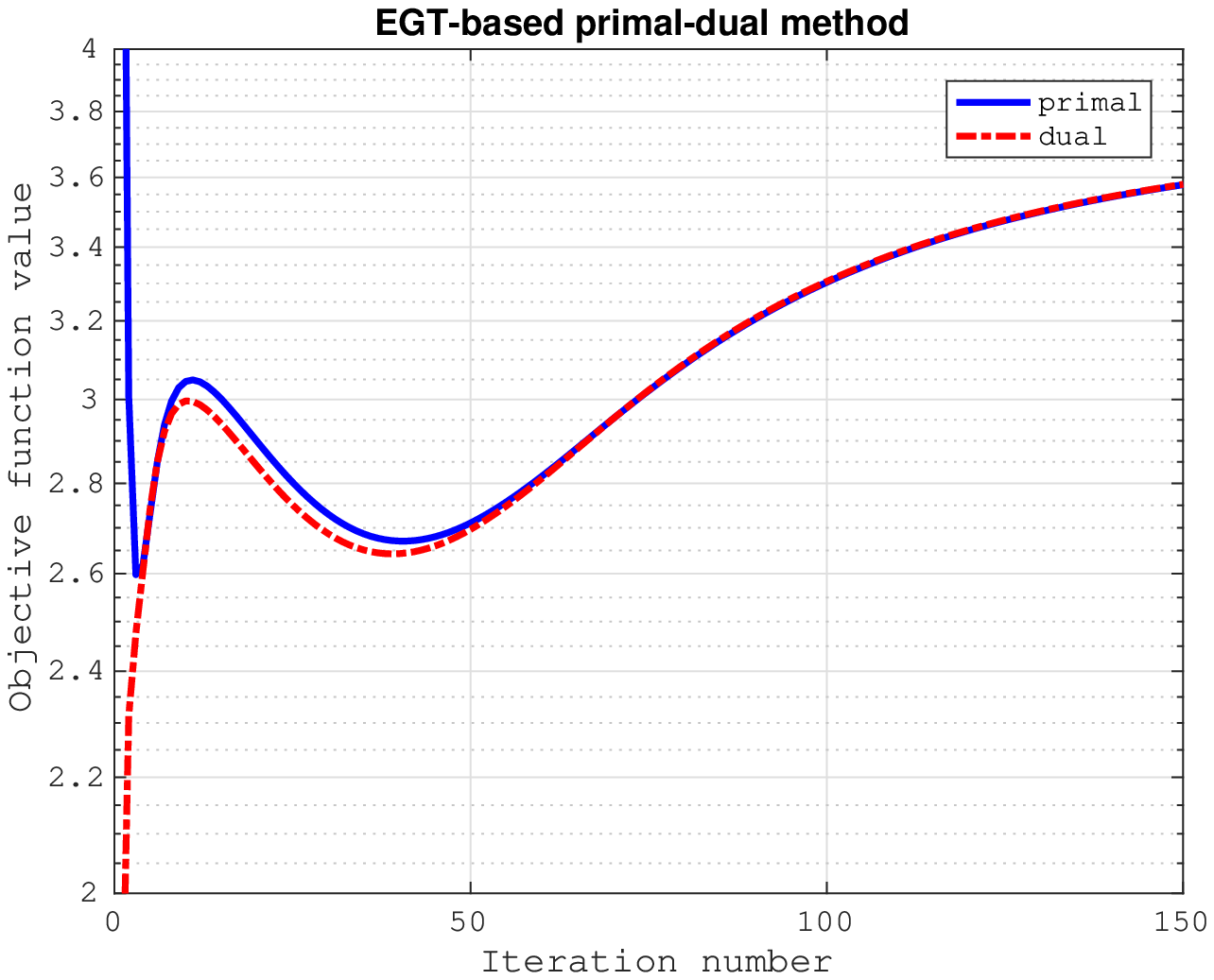}
\caption{Convergence of the EGT-based primal-dual method at SNR = 0 dB.}\label{fig:EGT_primal-dual_SNR0}
\end{center}
\end{figure}

\begin{figure}
\begin{center}
\includegraphics[width=\columnwidth]{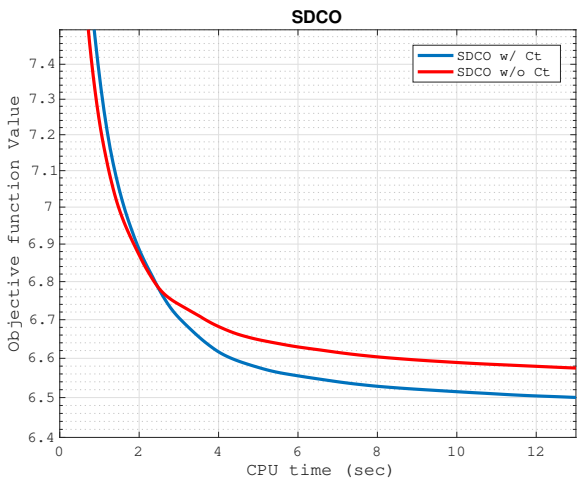}
\caption{Convergence comparison of SDCO with and without continuation at SNR=2 dB.}\label{fig:obj_val_snr2_v1}
\end{center}
\end{figure}

\begin{figure}[htb]
\begin{center}
\includegraphics[width=\columnwidth]{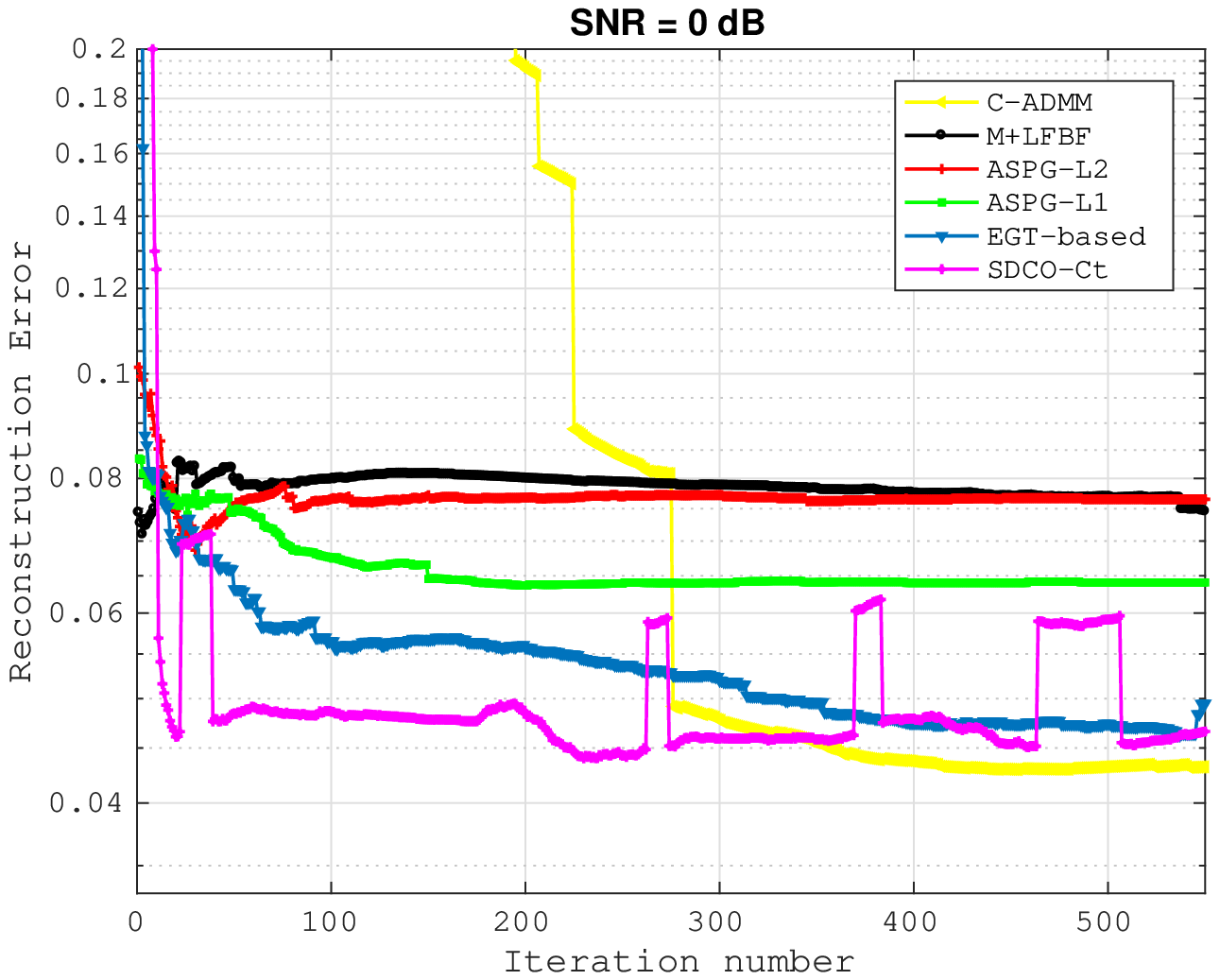}
\caption{Reconstruction error versus iteration number at SNR = 0 dB.}\label{fig:reconstr_err_SNR0}
\end{center}
\end{figure}

\begin{figure}[htb]
\begin{center}
\includegraphics[width=\columnwidth]{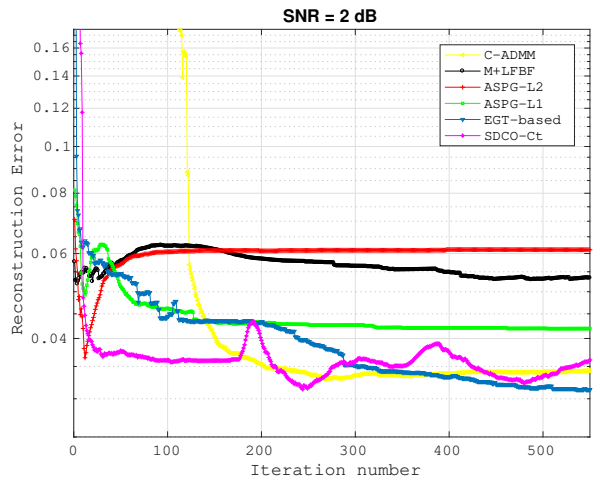}
\caption{Reconstruction error versus iteration number at SNR = 2 dB.}\label{fig:reconstr_err_SNR2}
\end{center}
\end{figure}

\subsection{Convergence Performance Comparisons}
The convergence performance of the proposed methods is verified in this numerical simulation in terms of reconstruction error or objective function value. The reconstruction error is defined as $E[ \frac{\| \hat{{\boldsymbol \theta}} - {\boldsymbol \theta} \|_2}{\| {\boldsymbol \theta} \|_2} ]$. First, we inspect  the convergence of the EGT-based method, in which the smoothing parameters for primal and dual problem are chosen with respect to iteration numbers, which is like the diminishing step size rule \cite{bertsekas1999nonlinear}. As shown in Figure \ref{fig:EGT_primal-dual_SNR0}, the duality gap becomes very small after the iteration number achieves 50. Second, the convergence comparison between the SDCO with and without continuation is conducted.  In Figure \ref{fig:obj_val_snr2_v1}, the convergence rate of the SDCO with continuation is almost the same as the one without continuation. However, it can achieves a lower objective function value that leads to better accuracy performance, since the smoothing parameter is reduced gradually by the continuation technique. \\
Finally, we inspect the convergence performance of C-ADMM, M+LFBF, ASPG-L1, ASPG-L2, EGT-based, and SDCO-Ct. In Figure \ref{fig:reconstr_err_SNR0}, at SNR $=0$ dB,  M+LFBF, ASPG-L1, ASPG-L2, EGT-based, and SDCO-Ct converge after iteration number is 100, while C-ADMM converges after iteration number is 300. Only SDCO-Ct, EGT-based, and C-ADMM can have lowest reconstruction error among them, but SDCO-Ct seems unstable in this case.  In Figure \ref{fig:reconstr_err_SNR2}, at SNR $=2$ dB, the convergence rate of C-ADMM gets improved., but is still slower than all the others. The SDCO-Ct method is the fastest one to converge to the lowest reconstruction error, and the unstableness is much less than the previous case.

\section{Conclusion}
In this paper, several iterative methods with the Nesterov smoothing technique were proposed for the estimation of off-grid DoAs.  First, the C-ADMM method is applied. In order to improve the convergence rate of C-ADMM, two reformulation of the group-sparsity penalty is introduced and smoothed by the Nesterov smoothing technique so that its gradient can be calculated easily. Then, the accelerated proximal gradient is used to solve the unconstrained optimization problem with the smoothed objective functions plus the nonsmooth indicator function. The smoothing parameter is selected empirically. Thus, the variant of EGT-based method is employed so that the smoothing parameter can be chosen systematically. Instead of heuristically choosing a regularization parameter in the BPDN problem formulation, the variant of SDCO method is proposed, and its smoothing parameter can also be decided by using the continuation technique.
The accuracy performance and convergence of the proposed methods were verified by a numerical example of DoA estimation.


\appendices

\section{Proof of Lemma \ref{conv_anal}}

 \begin{proof}
Denote the smoothed version of the objective function $F({\bf x})$ as
\begin{align}  
\min_{{\bf x} \in {\mathbb R}^n}    F^{l_i}({\bf x})=\{f({\bf x})+h_{\mu}^{l_i}({\bf x})+ \iota_{\mathcal X}({\bf x}) \},  i=1 \text{~or~} 2 
\end{align} 
with the Lipschitz continuous gradient constant $L=L_f+\frac{1}{\mu \sigma}$.
By using similar proof schemes in \cite{lan2011primal}, we decompose 
\begin{align}  \nonumber
 F({\bf x}^k) -  F({\bf x}^*) =& (  F({\bf x}^k) -  F^{l_i}({\bf x}^k)  )+ \\
&(  F^{l_i}({\bf x}^k) -  F^{l_i}({\bf x}^*)   ) + (  F^{l_i}({\bf x}^*) -  F({\bf x}^*)  ).
\end{align} 
Then, based on the theorem from \cite{beck2009fast}, we have the following bound for an optimal solution $\bf x^*$:
\begin{align} 
F({\bf x}^k)-F({\bf x}^*) \leq \frac{2 L_f \|  {\bf x}^0 - {\bf x}^* \|^2}{(k+1)^2}.
\end{align} 
Also, by the definition of $h_{\mu}^{l_i}({\bf x})$, we have
\begin{align}  
  F^{l_i}({\bf x}^k)  \leq  F({\bf x}^k)  \leq  F^{l_i}({\bf x}^k)   + \mu D_i.
\end{align} 
This implies that 
\begin{align}  
&  F({\bf x}^k) -  F^{l_i}({\bf x}^k)   \leq \mu D_i.  \\
&  F^{l_i}({\bf x}^*) - F({\bf x}^*)  \leq 0. 
\end{align} 
Thus, 
\begin{align}  
 F({\bf x}^k) -  F({\bf x}^*)   & \leq \mu D_i +     \frac{2 L  \|  {\bf x}^0 - {\bf x}^* \|^2}{(k+1)^2} \\
&= \mu D_i +     \frac{2 (L_f+\frac{1}{\mu \sigma})\|  {\bf x}^0 - {\bf x}^* \|^2}{(k+1)^2}.
\end{align} 
Let $\mu = \frac{ \epsilon}{2D_i}$, then
\begin{align}  
 F({\bf x}^k) -  F({\bf x}^*)   \leq  \frac{\epsilon}{2}+     \frac{2 (L_f+\frac{2D_i}{\epsilon \sigma})\|  {\bf x}^0 - {\bf x}^* \|^2}{(k+1)^2}.
\end{align} 
If we let $ \frac{\epsilon}{2}+     \frac{2 (L_f+\frac{2D_i}{\epsilon \sigma})\|  {\bf x}^0 - {\bf x}^* \|^2}{(k+1)^2} = \epsilon$,
then we have the upper bound in (\ref{epsilon_appr}). 
\end{proof}


%
%
%
%
%
%

\ifCLASSOPTIONcaptionsoff
  \newpage
\fi



%

%
%
%

\bibliographystyle{IEEEtran}
\bibliography{IEEEabrv,mybib_ch1,mybib_ch2,mybib_ch3,mybib_ch5}

%

\begin{IEEEbiography}{Cheng-Yu Hung}
\end{IEEEbiography}



\begin{IEEEbiographynophoto}{Mostafa Kaveh}
\end{IEEEbiographynophoto}




\end{document}